\documentclass[prd,aps,amsmath,twocolumn,superscriptaddress,amssymb,reprint,nofootinbib,preprintnumbers,longbibliography]{revtex4-2}

\usepackage{graphicx,array}
\usepackage[hidelinks]{hyperref}
\usepackage{color}
\usepackage[T1]{fontenc}
\usepackage{amsmath,amssymb,slashed,latexsym}
\usepackage{bm}
\definecolor{lightblue}{rgb}{0.1, 0.5, 1.0}

\DeclareRobustCommand{\Sec}[1]{Sec.~\ref{#1}}
\DeclareRobustCommand{\Secs}[2]{Secs.~\ref{#1} and \ref{#2}}

\DeclareRobustCommand{\App}[1]{App.~\ref{#1}}

\DeclareRobustCommand{\Fig}[1]{Fig.~\ref{#1}}

\DeclareRobustCommand{\Eq}[1]{Eq.~(\ref{#1})}
\DeclareRobustCommand{\Eqs}[2]{Eqs.~(\ref{#1}) and (\ref{#2})}

\DeclareRobustCommand{\Reff}[1]{Ref.~\cite{#1}}
\DeclareRobustCommand{\Reffs}[1]{Refs.~\cite{#1}}

\hypersetup{
    colorlinks=true,
    linkcolor=blue,
    filecolor=magenta,      
    urlcolor=blue,
    pdftitle={wifi ensembles},
    pdfpagemode=FullScreen,
    citecolor=blue
    }
    
\begin{document}

\preprint{MIT-CTP/5874}

\title{\texorpdfstring{\boldmath Frequentist Uncertainties on Neural Density Ratios with $w_i f_i$ Ensembles}{Frequentist Uncertainties on Neural Density Ratios with wifi Ensembles}}

\author{Sean Benevedes}\email{seanmb@mit.edu}
\author{Jesse Thaler}\email{jthaler@mit.edu}
\affiliation{Center for Theoretical Physics -- a Leinweber Institute, Massachusetts Institute of Technology,\\
Cambridge, Massachusetts, United States}
\affiliation{The NSF AI Institute for Artificial Intelligence and Fundamental Interactions}

\begin{abstract}
We introduce $w_i f_i$ ensembles as a novel framework to obtain asymptotic frequentist uncertainties on density ratios, with a particular focus on neural ratio estimation in the context of high-energy physics.
When the density ratio of interest is a likelihood ratio conditioned on parameters, $w_i f_i$ ensembles can be used to perform simulation-based inference on those parameters.
After training the basis functions $f_i(x)$, uncertainties on the weights $w_i$ can be straightforwardly propagated to the estimated parameters without requiring extraneous bootstraps.
To demonstrate this approach, we present an application in quantum chromodynamics at the Large Hadron Collider, using $w_i f_i$ ensembles to estimate the likelihood ratio between generated quark and gluon jets.
We use this learned likelihood ratio to estimate the quark fraction in a synthetic mixed quark/gluon sample, showing that the resultant uncertainties empirically satisfy the desired coverage properties. 
\end{abstract}

\maketitle
\flushbottom
{\small
\tableofcontents
}


\section{Introduction}
\label{sec:intro}

Machine learning is used ubiquitously in high-energy physics (HEP) for a wide breadth of applications, including jet tagging~\cite{CMS:2017wtu, Kasieczka:2019dbj, Bols:2020bkb, Qu:2022mxj,Gong:2022lye,Bogatskiy:2023nnw}, anomaly detection~\cite{ATLAS:2020iwa,Kasieczka:2021xcg,Aarrestad:2021oeb,Metodiev:2023izu,ATLAS:2023azi,ATLAS:2023ixc,CMS:2024nsz,Gambhir:2025afb}, supervised searches for new physics~\cite{ATLAS:2019vwv,ATLAS:2020pcy,ATLAS:2020tlo,CMS:2021yci}, unfolding~\cite{Andreassen:2019cjw,Bellagente:2019uyp,Bellagente:2020piv,Vandegar:2020yvw,Andreassen:2021zzk,Arratia:2021otl,Howard:2021pos,Backes:2022sph,Shmakov:2023kjj,Shmakov:2024gkd}, and generative modeling~\cite{Otten:2019hhl,Hashemi:2019fkn,DiSipio:2019imz,Butter:2019cae,Alanazi:2020klf,Butter:2020tvl,Butter:2021csz,Buhmann:2023pmh,Krause:2024avx};
see \Reff{Feickert:2021ajf} for a more comprehensive bibliography.
At the heart of many of these use cases is density ratio estimation (DRE), which is the task of inferring the ratio of two probability densities given samples from each density but not the functional form of the densities themselves.
A paradigmatic example of DRE in HEP is for classification tasks like jet tagging: according to the Neyman-Pearson lemma \cite{Neyman:1933wgr}, the optimal binary classifier is given by any monotonic function of the density ratio of the two likelihoods.
The ratio of two densities can also be used to reweight samples from one density to obtain samples from the other, as is relevant for detector unfolding (see e.g.~\cite{Andreassen:2019cjw}).
The primary application we consider here is for parameter estimation, where DRE is the foundation for various techniques in simulation-based inference (SBI)~\cite{Cranmer:2015bka}.

In practice, DRE is performed with a finite amount of data and compute, so the resultant point estimate of the density ratio will not be exactly equal to the true value.
This discrepancy should be quantified as an uncertainty and propagated to downstream applications; see \Reff{hermans2022} for case studies where parameter estimation is unreliable due to mismodeling of the density ratio.
To the best of our knowledge, there are no existing frequentist methods that rigorously assess uncertainties on neural density ratios directly.
Rather, the state of the art for SBI in HEP is to perform the Neyman construction~\cite{neymanconstruction}, where one treats the estimated density ratio as an arbitrary test statistic and uses bootstrapping to determine its empirical distribution.
While the Neyman construction produces statistically rigorous uncertainties for downstream parameter inference, it is computationally expensive due to the requisite bootstrapping, and it is not applicable when one requires a notion of uncertainty on the density ratio itself rather than on a downstream test statistic.

In this paper, we introduce $w_i f_i$ ensembles, a framework for DRE that uses model ensembles to produce statistically rigorous point estimates of, and uncertainties on, density ratios.
To estimate a density ratio $r(x)$ of two distributions $n(x)$ and $d(x)$,
\begin{equation}
\label{eq:r_def}
    r(x) \equiv \frac{n(x)}{d(x)},
\end{equation}
we propose the model
\begin{equation}
    \label{eq:intro_wifi}
    \log \tilde{r}(x|w) = \sum_i w_i f_i(x),
\end{equation}
where $\tilde{r}(x|w)$ is our parametrized model of $r(x)$, the $f_i(x)$ are neural networks (NNs) used to model the density ratio, and the $w_i$ are weights inferred from training data (not to be confused with NN weights).
We use results on $M$-estimators~\cite{huber64} to deduce asymptotic coverage properties for the confidence intervals of the $w_i$ (and by extension, for confidence intervals on the density ratio itself).
We also show how these uncertainties on the density ratio can be propagated using the Gong-Samaniego theorem~\cite{gstheorem} to obtain downstream uncertainties on parameters in the context of SBI.
Our results rely on two assumptions, both of which amount to conditions on the training of the $f_i(x)$.
The first assumption is that the model is \textit{well-specified}: that there exists a set of weights $w^*_i$ for which $\tilde{r}(x|w^*) = r(x)$.
The second assumption is that we can treat the $f_i(x)$ as fixed for the purposes of fitting the $w_i$. 
This is necessary in principle for our formulae (which do not include correlations between the $f_i(x)$ and the $w_i$) to be applicable, although empirically we have not found any evidence for such correlations.
In practice, this can be achieved by using separate datasets for the purposes of training the $f_i(x)$ and fitting the $w_i$.
Intuitively, the ensemble weights $w_i$ play the role of nuisance parameters in a collider physics analysis.
If we only use a point estimate of a density ratio, e.g.\ directly from conventional NN training, then there is a resultant unquantifiable mismodeling of the density ratio (a systematic, in the collider physics parlance).
Introducing and fitting the weights $w_i$ trades off this unquantifiable mismodeling for an additional, quantifiable source of uncertainty (i.e.\ a statistical uncertainty).
Note that $w_i f_i$ ensembles cannot solve the general challenge of model selection, and our approach is only as reliable as the assumed model form in \Eq{eq:intro_wifi}.

The distinction between mismodeling and uncertainty is a crucial point:
\begin{itemize}
    \item \emph{Mismodeling} (also known as model misspecification) is a mismatch between the true data-generating process and a particular model of that process.
    In the case of a parametrized model, mismodeling occurs when there is no set of parameters that reproduce the data-generating distribution.
    Mismodeling is generally unquantifiable, and collecting more data does not help.
    \item \emph{Uncertainty} is the remaining degree of ignorance after performing statistical inference, assuming the absence of mismodeling.
    Uncertainty is generally quantifiable and systematically improvable by collecting more data.
\end{itemize}
Factors such as limited NN expressivity as well as stochasticity of NN initialization and training therefore potentially contribute to mismodeling, but are not uncertainties in this framing.
The aim of $w_i f_i$ ensembles is to trade off the unquantifiable mismodeling due to limited training data and training stochasticity in favor of quantifiable uncertainties.
They do this by introducing a parametrized model which is more flexible than each individual basis function $f_i(x)$, and then quantifying the uncertainties through the introduced weights $w_i$.
To demonstrate our approach, we focus on one particular use case for DRE: parameter estimation.
Parameter estimation is a central task in HEP, and a tremendous amount of work has gone into precise determinations of parameters like the strong-coupling constant $\alpha_s$~\cite{ParticleDataGroup:2024cfk,Salam:2017qdl,Pich:2018lmu,Proceedings:2019pra, Pich:2020gzz, dEnterria:2022hzv} and the top quark mass $m_t$~\cite{Nason:2017cxd, Hoang:2020iah, ATLAS:2024dxp}.
Traditionally, these parameters are measured by collecting a large amount of high-dimensional data (e.g.\ collider events), reducing the dimensionality of the data by projecting onto a small number of handcrafted observables, and binning to estimate the likelihood conditioned on the parameter(s) of interest.
However, these processing steps throw away some of the information in the data.
SBI, using DRE, provides a means of fully exploiting this information by estimating the (full, high-dimensional) likelihood ratio.
Our case studies will demonstrate how $w_i f_i$ ensembles can streamline the SBI process by quantifying the uncertainty on inferred parameters through an uncertainty estimate on the likelihood ratio directly.

Existing methods that use ensembles of NNs for uncertainty quantification, like deep ensembles~\cite{lakshminarayanan2017simplescalablepredictiveuncertainty}, combine the outputs of the NNs through an unweighted average.
In our case, the weights $w_i$ assigned to each of the models in a $w_i f_i$ ensemble are free parameters determined from the training data.
This allows us to formalize the idea that not all members of an ensemble will be of equal utility, and so they should not contribute equally.
Separate from uncertainty quantification, NN ensembles are widely used to improve model robustness~\cite{10.1007/3-540-45014-9_1}.
When combined with the Neyman construction to quantify uncertainties, though, it can be computationally daunting to train a bootstrapped ensemble of NN ensembles~\cite{ATLAS:2025clx}.
In our case, because the weights $w_i$ can be treated just like nuisance parameters, we can use asymptotic formulae to obtain uncertainties without extraneous bootstraps.
Of course, $w_i f_i$ ensembles are still an ensemble, and we will advocate for bootstrapping as an effective way to construct that ensemble; the point to emphasize is that we only need \emph{one} ensemble to achieve both robustness and uncertainty quantification at the same time.

The remainder of the paper is organized as follows. 
In \Sec{sec:background}, we review DRE, its connection with SBI, and the necessity of uncertainties in the inferred density ratio.
Then, in \Sec{sec:method}, we introduce $w_i f_i$ ensembles in detail, explaining how the density ratio and its uncertainties are inferred from data, and how these uncertainties are propagated for parameter estimation.
In \Sec{sec:gaussian}, we show empirically that $w_i f_i$ ensembles successfully perform DRE with well-calibrated uncertainties in a Gaussian toy example where the density ratio is known. 
We then move to an application in the context of quantum chromodynamics (QCD) in \Sec{sec:qg}, showing that $w_i f_i$ ensembles can be used to estimate the quark fraction in a synthetic mixed sample of generated quark and gluon jets.
We conclude in \Sec{sec:conclusions} by discussing future directions.
Additional statistical details appear in the appendices:  \App{app:what} discusses the properties of our $w_i$ estimator, \App{app:plrt} describes an alternative pseudo-likelihood method for producing confidence intervals, and \App{app:bias} studies the leading bias of the $w_i$ estimator in the asymptotic limit.
Implementations of $w_i f_i$ ensembles, and the code used in our experiments, can be found at \url{https://github.com/benevedes/wifi-ensembles}.

\section{Review of Density Ratio Estimation}
\label{sec:background}

In this section, we review DRE and its applications to frequentist parameter estimation.
In \Sec{subsec:def}, we introduce the DRE task.
Then, in \Sec{subsec:param_est_rev}, we review how DRE can be used to estimate likelihood ratios, enabling SBI.
In \Sec{subsec:uncertainties}, we discuss the uncertainties and mismodelings that arise in DRE, and their effects in the context of SBI.
Finally, in \Sec{subsec:existing_work}, we describe existing frequentist approaches for estimating these uncertainties.

\subsection{Problem Statement}
\label{subsec:def}

Consider two probability distributions $n(x)$ and $d(x)$ of an arbitrary random variable $x$, which we call the numerator and denominator distributions.
Suppose further that it is possible to obtain samples drawn from each of these distributions, perhaps through experiment or through simulation, but that the functional forms of the distributions themselves are not known.
The DRE task is then, given a dataset $D$ consisting of samples drawn from each of the distributions, to estimate the ratio
\begin{equation}
r(x) \equiv \frac{n(x)}{d(x)}.
\end{equation}
Throughout, we take samples to be independent and identically distributed (i.i.d.).

\subsection{Parameter Estimation}
\label{subsec:param_est_rev}

Now, consider the parameter estimation task.
Suppose that an experiment is performed to measure some set of parameters $\Theta$.
The data consist of phase space variables $\Phi$ collected by the experiment that are sampled according to some probability distribution $p(\Phi|\Theta^*)$, where $\Theta^*$ denotes the set of ground truth values of the $\Theta$.
To perform frequentist inference on $\Theta$, we could perform a maximum likelihood estimate (MLE) analysis if we knew the family of distributions $p(\Phi|\Theta)$ as a function of $\Theta$.
However, it is not immediately obvious how to obtain the MLE in the situation where $p(\Phi|\Theta)$ is unknown, as is often the case in collider physics; it is intractable to directly calculate the probability density over collider events.

DRE provides a way forward in this situation, under the assumption that one can sample (i.e.\ through simulation) the joint distribution $p(\Phi,\Theta)$.%
\footnote{Note that this joint distribution implies the existence of a prior $p(\Theta)$.  The inference procedures that we consider are frequentist in the sense that they are independent of this prior so long as the prior has support over all realizable values of $\Theta$. We assume this support property throughout.}
The ability to sample from this joint distribution immediately implies that one can also sample from the product distribution $p(\Phi)\, p(\Theta)$; each of the marginals $p(\Phi)$ and $p(\Theta)$ can be sampled by sampling the joint and throwing away the information about the marginalized variables, and the product distribution can then be sampled by pairing a sample from $p(\Phi)$ with a sample from $p(\Theta)$.
Then, we can perform DRE with $x = \{\Phi, \Theta\}$ using $p(\Phi,\Theta)$ as the numerator distribution and $p(\Phi) \, p(\Theta)$ as the denominator distribution:
\begin{equation}
\label{eq:llr}
    \ell(\Phi|\Theta) \equiv \frac{p(\Phi, \Theta)}{p(\Phi) p(\Theta)} = \frac{p(\Phi|\Theta)}{p(\Phi)}
\end{equation}
This is then a frequentist procedure for so-called likelihood-free or simulation-based inference (SBI); this specific class of methods is called neural ratio estimation \cite{Cranmer:2015bka} if NNs are used to parametrize the estimated density ratio.

As opposed to traditional statistical methods in collider physics, which employ dimensionality reduction to a set of hand-selected summary statistics and binning to estimate likelihoods, SBI allows one to perform inference using the full phase space information of the data.
This, in principle, allows for more sensitive measurements and probes for new physics.

\subsection{Uncertainties in Density Ratio Estimation}
\label{subsec:uncertainties}

When we consider the DRE procedure in more detail, however, important complications arise.
Practically, one encounters various sources of \emph{mismodeling}, meaning a mismatch between
\begin{itemize}
    \item The true (but unknown) data-generating distribution as a function of parameters of interest $p(\Phi|\Theta)$, and
    \item The proxy distributions that we can access either through sampling or through evaluation as a function of data.
\end{itemize}
We similarly refer to a density ratio as being mismodeled if there is a mismatch between the ratio of the true data-generating distributions and the available proxy ratio.

We then use the term \emph{uncertainty} to refer to our remaining ignorance of the true parameters $\Theta^*$ in the absence of mismodeling.
Under mild assumptions like finite variance, uncertainty tends to zero as the amount of i.i.d.\ data collected tends to infinity, while errors induced by mismodeling do not.%
\footnote{Later, we refer to NN training as a potential source of mismodeling.  This mismodeling generically decreases as the amount of NN training data increases, which might at first glance seem contradictory with the previous statement.  The resolution of this tension is that, in the contexts we describe, the training procedure is separate from the inference procedure.
When we refer to mismodeling due to NN training, we have in mind mismodeling at inference time.
At that point, the NNs should be regarded as fixed, so this source of mismodeling does not decrease as the size of the inference dataset increases.} 
Moreover, when we produce confidence intervals, these address uncertainties but not mismodeling; their coverage guarantees apply only in the absence of mismodeling.
In HEP parlance, our usage of mismodeling corresponds to unquantified systematic uncertainties, and our usage of uncertainties corresponds to quantified statistical and systematic uncertainties.

The source of mismodeling that we aim to reduce in this paper (and replace with quantified uncertainties) arises from the DRE itself.
In particular, we do not address any mismodeling that could arise if the training data are imperfectly sampled from $n$ and $d$, i.e.\ mismodeling due to flawed simulation.
We consider DRE procedures which model the density ratio with NNs; these networks are trained with loss functions which, asymptotically, admit the density ratio as the objective function.
However, in practice, the NN only approximately learns this objective function.
This is due to a variety of factors: the finite amount of data that the NN is trained on does not contain enough information to uniquely identify the ground truth density ratio, the training procedure does not perfectly exploit the information that is in the data, and a fixed, finite-width NN architecture is generically not expressive enough to capture the true density ratio.

The learned density ratio $\hat{\ell}(\Phi| \Theta)$ is then only approximately equal to $\ell(\Phi |\Theta)$, and the estimator for $\Theta$ obtained via maximization of $\hat{\ell}(\Phi | \Theta)$ is not the MLE that would have been obtained from the ground truth $\ell(\Phi|\Theta)$.
Then, in general, this estimator does not enjoy the various nice properties of the MLE; in particular, it generally is not consistent (i.e.\ converge in probability to $\Theta^*$ as the number of samples drawn from $p(\Phi|\Theta^*)$ goes to infinity), and the calculation of its variance through asymptotic formulae is generically incorrect.%
\footnote{One can gain intuition for these effects in a toy example.  Consider a normal distribution with unknown ground truth mean $\mu$ and variance $\sigma^2$. Suppose a devious rival scientist tells you under the guise of collaboration that the true value of the mean is a false value $\mu_\text{wrong}$, and it only remains to measure $\sigma^2$. The MLE for the variance, $\hat{\sigma}^2 = \frac{1}{N}\sum_i(x_i - \mu_\text{wrong})^2$, is then biased: $E[\hat{\sigma}^2] = (\mu - \mu_\text{wrong})^2 + \sigma^{2}$. This arises because the normal distribution with mean $\mu_\text{wrong}$ mismodels the data, so downstream inference becomes unreliable. In this analogy, $\mu$ corresponds to the density ratio being estimated and $\sigma$ corresponds to a parameter of interest. $w_i f_i$ ensembles then address this mismodeling by estimating the density ratio with appropriate uncertainties, and propagating these uncertainties for inferring the parameter(s) of interest.}

These discrepancies between estimated and true density ratios can be ruinous for SBI.
As shown in \Reff{hermans2022}, existing methods for DRE can lead to underestimated uncertainties and overconfident intervals for inferred parameters.
Therefore, to use DRE for parameter estimation, one needs not only a point estimate of the density ratio of interest, but also a notion of uncertainty on the estimate.

\subsection{Existing Work}
\label{subsec:existing_work}

Various approaches have been proposed to estimate uncertainties on NN outputs, both in DRE and in other contexts.
The bulk of the machine learning literature has investigated Bayesian approaches with NNs: conceptually, instead of conventional NN training, one instead starts with a prior on network weights.
Training then corresponds to a Bayesian update to obtain the weight posteriors (see e.g.~\Reff{arbel2023} for a recent review of Bayesian methods).
Once these posteriors are known, uncertainties on the network outputs can be estimated by sampling weight configurations from the posterior.

In practice, due to the computational difficulties involved in exact Bayesian inference, a variety of methods are used to approximate these posteriors, like sampling methods~\cite{blundell2015} and variational inference~\cite{graves2011}. 
Various other techniques like Dropout~\cite{JMLR:v15:srivastava14a,pmlr-v48-gal16} and Repulsive Ensembles~\cite{dangelo2023repulsivedeepensemblesbayesian} can also be motivated as performing approximate Bayesian inference.
These Bayesian techniques work well in many domains, but in the collider physics context, their prior dependence is undesirable.

Frequentist methods are less well-explored.
Inference through a direct MLE analysis on NN weights using asymptotic formulae is infeasible for various reasons, including the difficulty of finding the global maximum of the likelihood as a function of the weights, the computational expense in computing and inverting the high-dimensional Hessian of the likelihood with respect to the weights, and the fact that one is rarely actually in an asymptotic regime for NNs due to their massive number of parameters.

In the face of these difficulties, the state of the art in collider physics is to use pseudo-experiments to perform the Neyman construction~\cite{neymanconstruction} (see \Reff{ATLAS:2025clx} for a recent tour de force implementation of these methods).
For example, for SBI, one still performs DRE to obtain an approximate likelihood ratio test statistic.
If this test statistic were precisely the likelihood ratio, one could use asymptotic results like Wilks's Theorem \cite{wilks} to determine its distribution and perform a hypothesis test.
Since this is not the case, one instead computes the test statistic on a large number of simulations, each generated with a fixed, counterfactual value of $\Theta^*$, to determine this distribution pseudo-empirically.

The Neyman construction enjoys the significant advantage that it provides reliable frequentist uncertainties under very general conditions: the quality of the trained NN determines the size of the resultant confidence intervals but not their validity.
However, it is computationally expensive: the number of simulations required is generically quite large.
In particular, since it relies on having simulated samples which reasonably sample the possible values of $\Theta$, this method suffers from a curse of dimensionality as the dimension of $\Theta$ grows.
These difficulties can be improved, but not eliminated, using methods like those proposed in Refs.~\cite{dalmasso2020confidencesetshypothesistesting,dalmasso2024likelihoodfreefrequentistinferencebridging}.

Another framework, similar to $w_i f_i$ ensembles, was developed and explored for DRE in \Reff{sugiyama2008} in the context of kernel methods.
To the best of our knowledge, these methods have not previously been applied in collider physics or used with trained basis functions.
Our approach of $w_i f_i$ ensembles, on the other hand, use trained NNs for the $f_i$.
We also note that there are some interesting parallels between $w_i f_i$ ensembles and methods like AdaBoost \cite{FREUND1997119}, as it has been shown in \Reff{NIPS2001_71e09b16} that there are deep connections between AdaBoost and the MLE of the ensemble weights.

\section{Introducing \texorpdfstring{$w_i f_i$}{wifi} Ensembles}
\label{sec:method}

In order to quantify uncertainties on a density ratio $r(x)$, we introduce $w_i f_i$ ensembles: a framework that leverages ensembles of NNs to perform DRE with native uncertainties.
To model the density ratio $r(x)$, we introduce the model 
\begin{equation}
\label{eq:wifi}
    \log \tilde{r}(x|w) = w_i f_i(x),
\end{equation}
where the $w_i$ are learned weights to be inferred from data, the $f_i(x)$ are trained NNs, and we employ Einstein notation to suppress the sum over the index $i$ (which we take to run from $1$ to $M$, the size of the ensemble).

We will see that uncertainties on the $w_i$ can be assessed from data.
Under the assumptions that the model is well-specified, i.e.\ that the $f_i$ are chosen such that there exists a true set of weights $w_i^*$ for which $\tilde{r}(x|w^*) = r(x)$, and that the $f_i(x)$ are fixed for the purposes of fitting the $w_i$, these uncertainties then immediately correspond to uncertainties on the density ratio.
The well-specified assumption is, in effect, an assumption that the ansatz in \Eq{eq:wifi} for the log density ratio does not suffer from mismodeling.

It is worth emphasizing that, in general, reducing mismodeling and estimating uncertainties are conceptually distinct goals.
    In the context of $w_i f_i$ ensembles, though, these goals are linked, since the weights $w_i$ not only increase the flexibility of the model but also facilitate the extraction of confidence intervals.
   Of course, as mentioned in \Sec{subsec:uncertainties}, $w_i f_i$ ensembles cannot reduce mismodeling or estimate uncertainties associated with flaws in the training data.

DRE with $w_i f_i$ ensembles consists of the following steps, which provides a rough outline for the rest of this section:
\begin{enumerate}
    \item Train $M$ networks $f_i(x)$ on a training dataset $D_\text{train}$ (\Sec{subsec:nn_training}).
    \item Freeze the $f_i$ and infer the values of the best fit weights $\hat{w}_i$ on a distinct dataset $D_\text{fit}$ by using a suitable optimization objective (\Sec{subsec:wi_fit}).
    \item Compute the covariance matrix $C_{ij} \equiv \text{Cov}[\hat{w}_i, \hat{w}_j]$ using analytic formulae (\Sec{subsec:est_cij}).
    \item Propagate these uncertainties, either to the density ratio of interest itself or to downstream applications (\Sec{subsec:param_estimation}).
\end{enumerate}
The crux of $w_i f_i$ ensembles is in steps 2 and 3, while step 1 is relatively standard and step 4 is application specific.
Finally, in \Sec{subsec:general_mix_frac} we present the specific parameter estimation task, mixture fraction estimation, that we consider in the later empirical case studies.

\subsection{\texorpdfstring{Training the $f_i(x)$}{Training the fi}}
\label{subsec:nn_training}

The intuition behind $w_i f_i$ ensembles is to capture the uncertainty in density ratios by parametrizing them with the ansatz \Eq{eq:wifi}, using parametric methods to estimate the uncertainty on the $\hat{w}_i$, and then propagating these uncertainties to the density ratio itself (and to downstream applications).
Clearly, if the density ratio of interest is to be written in the form \Eq{eq:wifi}, an arbitrary set of $f_i$ does not suffice.

In particular, we emphasize that the asymptotic formulae which we derive are formally only applicable when the model is well-specified, where $\log r = w_i^* f_i$ for some $w_i^*$.
Our chief objective then is to obtain a basis of functions $f_i$ such that the vector space they span contains $\log r$.%
\footnote{In reality, it is exceptionally unlikely that this vector space actually contains $\log r$ for any realistic choices of training procedure and ensemble size $M$. For our inferences to be valid in practice, this condition is sufficient but not necessary; rather than requiring the total absence of mismodeling, it is in practice enough for the unquantifiable errors induced by mismodeling to be subdominant to the quantifiable uncertainties.}

The optimal strategy to build such a basis in practice is not obvious.
For our empirical case studies, we consider strategies where the $f_i$ are individually trained to estimate the density ratio of interest.%
\footnote{We also tried a strategy where one preferred network is trained with \Eq{eq:sample_symMLC} and the remaining networks are trained to be decorrelated from the preferred network and from each other. In our empirical testing, this strategy did not perform better than strategies which train the networks on equal footing, and the training was significantly more computationally intensive than the other methods.}
The two strategies that we consider are:
\begin{itemize}
    \item \textit{Partition}: Partition the training data into $M$ disjoint groups, and use each of these groups to train one of the $f_i$ with \Eq{eq:sample_symMLC}.
    \item \textit{Bootstrap}: Train each of the $f_i$ networks with \Eq{eq:sample_symMLC} on a bootstrapped resample of the training data.
\end{itemize}
The motivation behind this selection of strategies comes from the context of uniformly weighted ensembles where, as shown in \Reff{louppe2015}, correlations between ensemble members introduce unnecessarily large variances which do not go to zero as the size of the ensemble becomes large.
These results do not carry over to $w_i f_i$ ensembles directly, as the $\hat{w}_i$ are themselves dependent on the $f_i$, but they motivate us to consider diverse $f_i$.
In addition to being diverse, the $f_i$ should also be well-trained estimators of their objective functions.
To see this, consider that a simple sufficient condition for the model to be well-specified is just for one of the $f_i$ to be equal to the log density ratio.
Intuitively, then, it is more plausible for the $w_i f_i$ functional form to be a well-specified model of the log density ratio of interest if the $f_i$ are already individually close to learning this log density ratio.
This motivates using standard best practices to train the $f_i$, from using an architecture well-suited to the data to using well-tuned hyperparameters.

The Bootstrap and Partition protocols both reduce correlations between the networks by training them on different training datasets, but represent different tradeoffs between diversity and individual performance of the $f_i$.
Bootstrap allows each of the $f_i$ to be trained with more data than Partition, but Partition totally eliminates the correlations between the $f_i$ induced by the training data whereas Bootstrap only reduces it.
One can easily conceive of other strategies, e.g.\ Bootstrap but using subsamples rather than resampling the whole dataset to accelerate training.
We restrict ourselves to Partition and Bootstrap for simplicity, but it would be interesting to explore training methodologies more exhaustively.
For the case studies in \Secs{sec:gaussian}{sec:qg}, we will also consider a Naive Ensemble protocol as a baseline method for the sake of comparison.
For Naive Ensemble, the $f_i$ are trained identically to those of Bootstrap, but the $w_i$ are taken to be fixed to $1/M$ rather than fit to data.
This means that Naive Ensemble does not provide a notion of uncertainty on density ratios, and no such uncertainty is propagated to downstream parameter estimation.
In the large $M$ limit, Naive Ensembles converge to a dataset-dependent average estimator, which will not generically be equal to the true density ratio.

Throughout, we use $M$ to denote the number of trained $f_i$. However, we find empirically that it is also helpful to add one additional $f_{0}(x) = 1$; the corresponding $w_0$ then has the interpretation of satisfying a normalization constraint on the density ratio that we will encounter in \Eq{eq:norm}.
The index $i$ then runs over $M+1$ values. We include this additional $f_{0}$ in all experiments throughout, except for the Naive Ensemble protocol where the weights are fixed and not fit to data (alternatively, this is equivalent to fixing $w_{0} = 0$).

In all cases, we use one dataset $D_\text{train}$ to train the $f_i(x)$ and another dataset $D_\text{fit}$ to fit the $w_i$.
This is because our results for coverage on the $w_i$ (and therefore on density ratios and downstream inferences) assume that the $f_i(x)$ are fixed functions, and formally they cannot be treated as fixed if they are determined through a training procedure which is dependent on the data used to fit the $w_i$.
One way of seeing that the $f_i(x)$ must be fixed is to note that the true values of the weights $w_i^*$ are themselves dependent on the $f_i(x)$; therefore, one cannot even talk about frequentist coverage on the $w_i$ unless the $f_i(x)$ are fixed.
In principle, then, we expect that using the same dataset to train the $f_i(x)$ and to fit the $w_i$ would neglect correlations between the $f_i(x)$ and the $w_i$, and so could result in confidence intervals which undercover.
That said, we have performed experiments using the same dataset to train the $f_i(x)$ and fit the $w_i$, and we have not observed this effect in practice.

For the sake of simplicity, we take all of the $f_i$ to have identical architectures and hyperparameters.
This choice may be suboptimal, as it has been shown (e.g.\ in \Reff{wenzel:2021}) that varying hyperparameters of each member in traditional ensembles can lead to increased diversity and better performance in that context.
We leave exploration of such training strategies to future work.

\subsection{\texorpdfstring{Finding $\hat{w}_i$}{Finding wi}}
\label{subsec:wi_fit}

Suppose that a set of $M$ functions $f_i(x)$ has been obtained through one of the aforementioned methods and frozen.
To estimate the values of the $w_i$, we want some analogue of the MLE.
However, we must account for two constraints following from proper normalization of $n$ and $d$:
\begin{equation}
\label{eq:norm}
    1 = \int \text{d}x \, d(x) \, e^{w_i f_i(x)} = \int \text{d}x \,  n(x) \, e^{-w_i f_i(x)}.
\end{equation}
As has been shown in \Reff{Nachman:2021yvi}, normalization constraints of this kind can be imposed through a Lagrange multiplier, and solving for this Lagrange multiplier for one of the constraints yields the maximum likelihood classifier (MLC) loss first considered in \Reffs{DAgnolo:2018cun,DAgnolo:2019vbw}.

The maximum likelihood objective in conjunction with one of these normalization constraints (i.e.\ the MLC loss up to a minus sign) is already sufficient to yield the density ratio of interest asymptotically.
However, to respect the symmetry of the problem under which $n$ and $d$ are interchanged and $w_i f_i \rightarrow -w_i f_i$, we introduce the following symmetrized MLC optimization objective:
\begin{align}
    \label{eq:symMLC}
    \mathcal{L}_\text{symMLC} &= \int \text{d}x\bigg[- w_i f_i(x) n(x) + d(x) \left(e^{w_i f_i(x)} - 1\right)\bigg] \nonumber \\
    &+\int \text{d}x\bigg[w_i f_i(x) d(x) + n(x) \left(e^{-w_i f_i(x)} - 1\right)\bigg].
\end{align}
It can be checked that the density ratio of interest minimizes \Eq{eq:symMLC}; thus, as long as the model is well-specified so that there exists a $w_i^*$ for which $w_i^* f_i$ is precisely this density ratio, $w_i^*$ minimizes the loss.

Specializing to a sample of $N_n$ i.i.d.\ draws from $n$ and $N_d$ i.i.d.\ draws from $d$, this loss can be estimated on the dataset $D_\text{fit}$ as:
\begin{align}
    \label{eq:sample_symMLC}
        \mathcal{L}_\text{samp} &= \left \langle - w_i f_i(x) + \left( e^{-w_i f_i(x)} - 1\right)  \right \rangle_n \nonumber\\
    &+ \left\langle w_i f_i(x) + \left( e^{w_i f_i(x)} - 1\right) \right \rangle_d,
\end{align}
where 
\begin{equation}
    \langle \cdot \rangle_{n} \equiv \frac{1}{N_{n}}\sum_{x \sim n}^{N_{n}} \left( \cdot \right),
\end{equation}
and similarly for $d$.
$N_n$ and $N_d$ need not be the same, but they are taken to have the same asymptotic power counting.
We then define the $\hat{w}_i$ as:
\begin{equation}
\label{eq:what}
    \hat{w}_i \equiv \text{argmin}_{w_i} \mathcal{L}_\text{samp}.
\end{equation}
As we show in \App{app:what}, under the assumption that the model is well-specified so that there exists a $w_i^*$ for which $w_i^* f_i = \log r$, $\hat{w}_i$ is an asymptotically unbiased estimator for the $w_i^*$.
\Eq{eq:sample_symMLC} is convex in the $w_i$, so the optimization is tractable and can be performed using Newton's method, for example.

\subsection{\texorpdfstring{Estimating $C_{ij}$}{Estimating Cij}}
\label{subsec:est_cij}

As we discuss further in \App{app:what}, $\hat{w}_i$ is an $M$-estimator~\cite{huber64}, which is a class of estimators that arise when the objective function is a sample average.
Asymptotically, $M$-estimators are Gaussian distributed, and the covariance matrix $C_{ij}$ for the $\hat{w}_i$ distribution can be estimated by the sandwich formula:
\begin{equation}
\label{eq:sandwich}
    C^{ij} = V^{ik} U_{kl} V^{lj},
\end{equation}
where
\begin{equation}
    V_{ik} = \left.\frac{\partial^2\mathcal{L}_\text{samp}(w)}{\partial w_i \partial w_k} \right|_{w = w^*}
\end{equation}
is the Hessian matrix of the loss with respect to the $w_i$ evaluated at $w_i^*$, $V^{ij} = V_{ij}^{-1}$ is the matrix inverse, and
\begin{equation}
    U_{kl} = \left.\text{Cov}\left[\frac{\partial \mathcal{L_\text{samp}}}{\partial w_k},\frac{\partial \mathcal{L_\text{samp}}}{\partial w_l} \right]\right|_{w = w^*}
\end{equation}
is the covariance matrix of the analogue of the score vector.
The derivatives can be calculated analytically due to the known, simple dependence of the model (and therefore of the loss) on the $w_i$:
\begin{align}
    \frac{\partial \mathcal{L}_\text{samp}}{\partial w_i} &=  \langle a_i(x) \rangle_n+  \langle b_i(x) \rangle_d,\\
    \frac{\partial^2 \mathcal{L}_\text{samp}}{\partial w_i \partial w_j} &= \left\langle f_i(x) f_j(x) e^{-w_k f_k(x)} \right\rangle_n \nonumber\\
    &+ \left\langle f_i(x) f_j(x) e^{w_k f_k(x)}\right\rangle_d,
\end{align}
where we have introduced the notation
\begin{align}
    a_i(x) &= -f_i(x)\left(1+ e^{-w_j f_j(x)}\right), \\ 
    b_i(x) &= f_i(x)\left(1+ e^{w_j f_j(x)}\right).
\end{align}
$V_{ik}$ can then be consistently estimated at this asymptotic order with the substitution $w_i^* \rightarrow \hat{w_i}$.

$U_{kl}$ can also be estimated with the same substitution by using the assumption that the data is i.i.d., so the score vector decomposes into two i.i.d.\ sums of contributions.
The covariance of these sums can then be estimated using the empirical covariance of the summands.
In particular, we can estimate
\begin{align}
    U_{kl} \sim \frac{1}{N_n} &\big(\langle a_k(x) a_l(x) \rangle_n 
    -\langle a_k(x)\rangle_n\langle a_l(x)\rangle_n \big) \nonumber \\
    +\frac{1}{N_d}&\big(\langle b_k(x) b_l(x) \rangle_d - \langle b_k(x) \rangle_d \langle b_l(x) \rangle_d \big).
\end{align}

Therefore, asymptotically, the covariance matrix $C_{ij}$ can be consistently estimated from $D_\text{fit}$ using the sandwich estimator \Eq{eq:sandwich}.
The corresponding uncertainty in the density ratio, estimated as 
\begin{equation}
    \log \hat{r}(x|\hat{w}) = \hat{w}_i f_i(x),
\end{equation}
can then be read off as
\begin{equation}
\label{eq:dre_unc}
    \sigma^2(x) = f_i(x) C_{ij} f_j(x).
\end{equation}
It can be seen from the form of \Eq{eq:sandwich} that $C_{ij} \sim O(N^{-1})$, like in the case of the MLE, and this scaling then holds for \Eq{eq:dre_unc} as well.

\subsection{\texorpdfstring{Parameter Estimation with $w_i f_i$ Ensembles}{Parameter Estimation with wifi Ensembles}}
\label{subsec:param_estimation}

With uncertainties on the $\hat{w}_i$ and on the density ratio in hand, we now move to the final step: leveraging the inferred density ratio and its uncertainties to perform downstream parameter estimation.

Suppose we have performed an experiment and obtained a dataset $D_\text{exp}$ of samples drawn from $p(\Phi|\Theta^*)$, where $\Theta^*$ is the true value of $\Theta$ to be inferred from the data and $\Phi$ are the observed phase space variables.
Suppose also that we have used the methods described above with $x = \{\Phi, \Theta\}$ to model $\log \ell(\Phi | \Theta) \equiv \frac{p(\Phi|\Theta)}{p(\Phi)}$ as $\log \tilde{\ell}(\Phi|\Theta,w) = w_i f_i(\Phi,\Theta)$, and we have obtained an estimate $\log \hat{\ell}(\Phi|\Theta,\hat{w}) \equiv \log \tilde{\ell}(\Phi|\Theta,\hat{w}) = \hat{w}_i f_i(\Phi,\Theta)$.

If it were the case that $\log \hat{\ell} = \log \ell$ exactly, then the estimator $\hat{\Theta}$ defined as
\begin{equation}
\label{eq:pseudoMLE}
    \hat{\Theta} \equiv \text{argmax}_\Theta \log\hat{\ell}(\{\Phi\}|\Theta,\hat{w})
\end{equation}
would be the MLE for the $\Theta$ (where we have used $\{\Phi\}$ to denote that $\log \hat{\ell}$ is to be evaluated on the whole dataset, i.e.\ by summing over i.i.d.\ data).
However, since it is generically the case that $\log \hat{\ell} \neq \log \ell$, $\hat{\Theta}$ is not actually the MLE, and asymptotic formulae for the variance of the MLE underestimate the variance of $\hat{\Theta}$ due to neglecting the stochasticity of $\log \hat{\ell}$ itself.

We can account for this additional uncertainty using the Gong-Samaniego theorem~\cite{gstheorem}.
In the terminology of Gong and Samaniego, $\hat{\ell}$ is a pseudolikelihood and $\hat{\Theta}$ is a pseudo-MLE.
Suppose for simplicity that $\Theta$ is one-dimensional, i.e.\ that there is one parameter of interest $\theta$.
Then, under mild regularity conditions about the existence and boundedness of derivatives of $\log \tilde{\ell}(\{\Phi\}|\theta, w)$, they show that $\hat{\theta}$ is asymptotically normally distributed with mean $\theta^*$ and variance $\sigma^2_\text{GS}$, which can be consistently estimated by%
\footnote{The usual statement of the Gong-Samaniego theorem includes an extra term proportional to the covariance of $\log \ell(\{\Phi\}| \theta^*, w^*)$ with respect to $\hat{w}_i$. Since the former depends only on the inference data $D_\text{exp}$ and the latter depends only on the data $D_\text{fit}$ with which $\hat{w}_i$ was determined, this covariance is $0$ in our case.}
\begin{equation}
    \label{eq:GS}
    \sigma^2_{\text{GS}} = \sigma^2_\text{MLE} \left(1 + \sigma^2_\text{MLE} A_i C_{ij} A_j\right),
\end{equation}
where
\begin{align}
\label{eq:derivatives}
    \sigma^2_{\text{MLE}} &\equiv -\left.\frac{\partial^2 \log \tilde{\ell}(\{\Phi\}|\theta,\hat{w})}{\partial^2\theta}\right|_{\theta=\hat{\theta}}^{-1},\\
    A_i &\equiv \left.\frac{\partial^2 \log \tilde{\ell}(\{\Phi\}|\theta, w)}{\partial \theta \partial w_i}\right|_{w=\hat{w},\, \theta=\hat{\theta}},
\end{align}
and $C_{ij}$ is, as before, the covariance of the weights which can be estimated from $D_\text{fit}$.
Putting the pieces together, to perform SBI with $w_i f_i$ ensembles, one must first train a set of neural networks $f_i$ with the dataset $D_\text{train}$, use the dataset $D_\text{fit}$ to infer the weights $\hat{w_i}$ using \Eq{eq:what}, and then estimate the covariance matrix $C_{ij}$ with \Eq{eq:sandwich}.
The estimated density ratio and uncertainties can then be used to obtain the pseudo-MLE $\hat{\theta}$ on $D_\text{exp}$, and to estimate its uncertainty $\sigma^2_\text{GS}$ using \Eq{eq:GS}.
As long as the choice of the $f_i$ is such that the mismodeling of the $w_i f_i$ parametrization of the density ratio is negligible, this procedure then yields asymptotically valid frequentist confidence intervals (or regions, in the multivariate case) for $\theta$.

\subsection{Mixture Fraction Estimation}
\label{subsec:general_mix_frac}

The algorithms presented above provide a general framework to perform SBI with $w_i f_i$ ensembles.
The case studies we consider in \Sec{sec:gaussian} and \Sec{sec:qg} take a special form, though, where the likelihood $p(\Phi|\theta)$ depends in a known, analytic way on the parameter $\theta$.
Dividing by an appropriate reference distribution, we can write a likelihood ratio suitable for an MLE of $\theta$ as
\begin{equation}
    L(\Phi|\theta) \equiv \frac{p(\Phi|\theta)}{p_\text{ref}(\Phi)} = F(r(\Phi), \theta),
\end{equation}
where $F$ is a function that we know analytically, and $r(\Phi)$ is a density ratio to be estimated from data but which only depends on $\Phi$ and not on $\theta$.
Modeling $r(\Phi)$ as $\tilde{r}(\Phi|w)$, and defining $\tilde{L}(\Phi|\theta,w) = F(\tilde{r}(\Phi|w), \theta)$, we can still perform inference for $\theta$ using \Eqs{eq:pseudoMLE}{eq:GS} under the substitution $\tilde{\ell}\rightarrow \tilde{L}$.
This approach allows us to take the derivatives defined in \Eq{eq:derivatives} analytically, without recourse to autodifferentiation or methods like the Gaussian Ansatz proposed in \Reffs{Gambhir:2022dut,Gambhir:2022gua}.

Concretely, consider the so-called \textit{mixture fraction task}: estimate the parameter $\kappa$ given a dataset $D_\text{mix}$ of size $N_\text{mix}$ drawn i.i.d.\ from the mixture model:  
\begin{equation}
    p(x|\kappa) = \kappa\,  n(x) + (1-\kappa)\, d(x),
\end{equation}
where $\kappa$ is  the mixture fraction and takes values between $0$ and $1$, and $n$ and $d$ are two probability distributions.
Then, since 
\begin{equation}
    \log L(x|\kappa) \equiv\log \frac{p(x|\kappa)}{d(x)} = \log \big(\kappa \, r(x) + (1-\kappa)\big),
\end{equation}
with $p_\text{ref}(x)$ taken to be $d(x)$, knowledge of the ratio $r(x)$ (again defined as $n/d$ according to \Eq{eq:r_def}) is sufficient to extract an MLE for $\kappa$.
Note that in this setup, $x = \Phi$ does \emph{not} depend on the parameter $\kappa$.

As we saw in \Sec{subsec:param_estimation}, a consistent estimator $\hat{r}(x|\hat{w})$ dependent on parameters $w_i$ can then be used in place of $r(x)$ at the cost of introducing the additional variance captured by \Eq{eq:GS}.
Since the dependence of the likelihood on $\kappa$ and $w_i$ takes a known analytic form, the requisite (estimates of) derivatives of the likelihood can be calculated analytically when $\log \hat{r} = \hat{w}_i f_i$ is a $w_i f_i$ ensemble:
\begin{align}
\label{eq:mix_frac_derivs}
    \sigma^2_{\text{MLE}} &= \left(\sum_{a=1}^{N_\text{mix}} \left(\frac{e^{\hat{w}_i f_i(x_a)} - 1}{\hat{\kappa} e^{\hat{w}_j f_j(x_a)} + (1-\hat{\kappa})}\right)^2\right)^{-1}, \nonumber \\
    A_i &= \sum_{a=1}^{N_\text{mix}} \frac{f_i(x_a) e^{\hat{w}_j f_j(x_a)}}{\left(\hat{\kappa} e^{\hat{w}_k f_k(x_a)} + (1-\hat{\kappa})\right)^2},
\end{align}
where the sum is over elements of $D_\text{mix}$.
Here, $\hat{\kappa}$ is the pseudo-MLE for $\kappa$ obtained from \Eq{eq:pseudoMLE} after plugging in the estimates $\hat{w}$: 
\begin{equation}
\label{eq:mf_pseudomle}
    \hat{\kappa} = \text{argmax}_\kappa \sum_{a=1}^{N_\text{mix}} \log \left( \kappa e^{\hat{w}_i f_i(x_a)} + (1-\kappa)\right).
\end{equation}
We empirically confirm the validity of these formulae in the following
case studies.

\section{Gaussian Example}
\label{sec:gaussian}

In this section, we show how $w_i f_i$ ensembles can be used to solve a one-dimensional Gaussian toy problem.
This simple problem, where the density ratio of interest is known exactly, allows us to check the coverage properties of $w_i f_i$ ensembles both for the density ratio itself and for a downstream SBI task.
In \Sec{subsec:gaussian_setup}, we describe the Gaussian toy problem in detail and establish the training methodology for our networks.
In \Sec{subsec:cov_test_method}, we discuss how we evaluate coverage of confidence intervals.
Then, in \Sec{subsec:gaussian_llr_results}, we present results for coverage on the density ratio itself.
Finally, in \Sec{subsec:gaussian_mf_results}, we examine coverage on an estimated parameter in the context of the mixture fraction task.

\subsection{Training Methodology}
\label{subsec:gaussian_setup}

Let the numerator distribution $n$ be a one-dimensional Gaussian with mean $\mu$ and variance $1$, and the denominator distribution $d$ also be a Gaussian with the same variance but mean $-\mu$.
The DRE task is then to estimate the log likelihood ratio of $n$ and $d$, $\log r(x) = 2 \mu x$.
We also consider the mixture fraction task, introduced in detail in \Sec{subsec:general_mix_frac}, to estimate the fraction $\kappa$ of events drawn from $n$ in a mixed population of events from $n$ and $d$.
The mixture fraction task gives us a window into the performance of $w_i f_i$ ensembles that does not require a known analytic form for the density ratio of interest, as relevant for the QCD case study considered in \Sec{sec:qg}.

We take the training sets $D_n$ and $D_d$ to each contain $N_n = N_d \equiv N = \text{25,000}$ events, and we choose $\mu = 0.1$.
We take the $f_i(x)$ to be feedforward NNs with one hidden layer and width $32$, using Leaky ReLU activation functions with leakiness $0.2$.
We find that this simple architecture is sufficiently expressive for this Gaussian toy problem.
To train the $f_i$, we use the Adam \cite{kingma2017adammethodstochasticoptimization} optimizer and early stopping with a patience of $10$ with a validation dataset, also with $N$ samples each from $n$ and $d$.
Throughout, we use PyTorch \cite{paszke2019pytorchimperativestylehighperformance} to implement and train NNs.

\subsection{Coverage Test Methodology}
\label{subsec:cov_test_method}

\begin{figure*}
    \centering
    \includegraphics[width=0.9\linewidth]{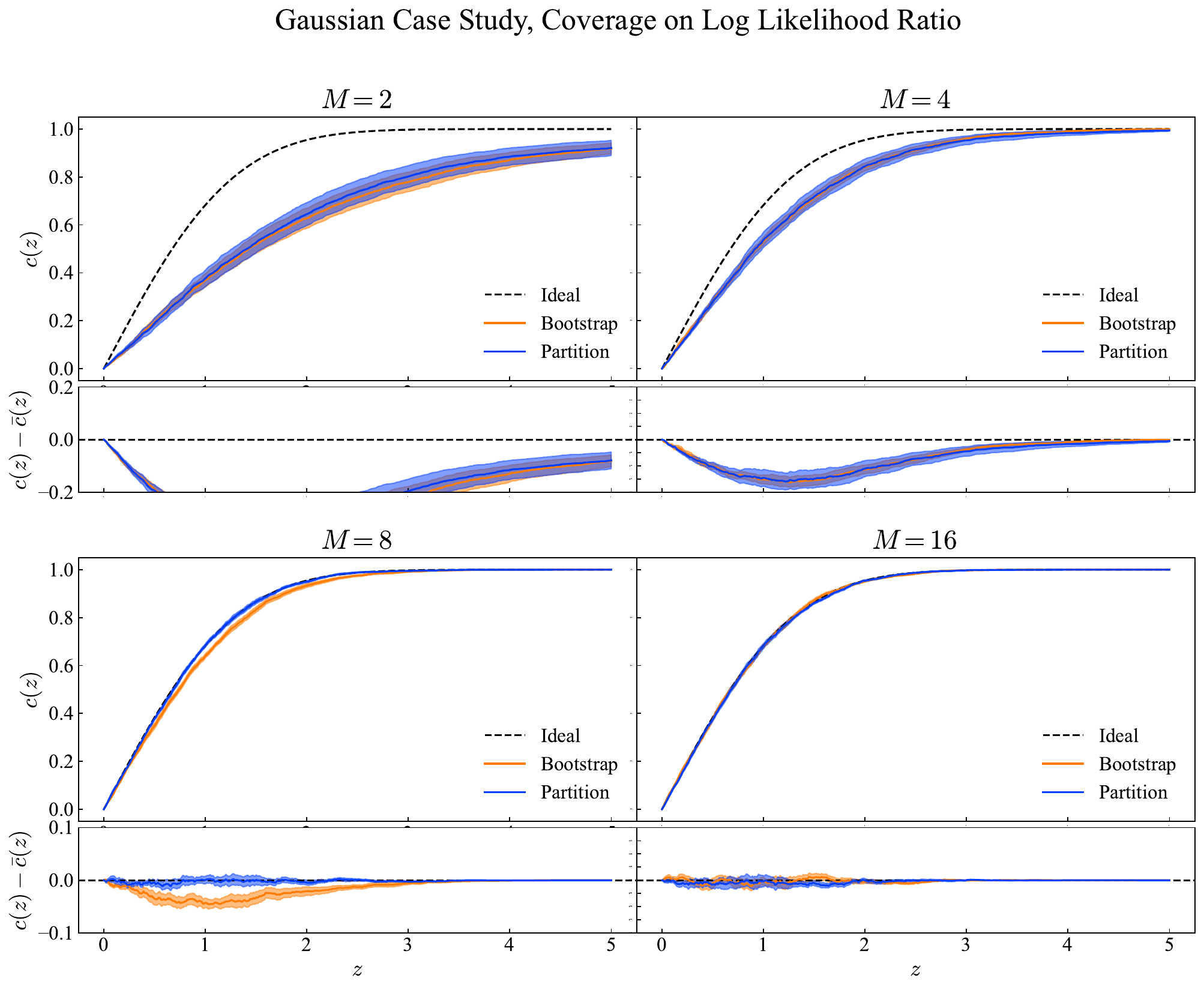}
    \caption{
    For the Gaussian case study, coverage $c(z)$ of a $z \sigma$ confidence interval for the density ratio $\log r$ for the Partition and Bootstrap training protocols from \Sec{subsec:nn_training}.
    Each quadrant corresponds to a different ensemble size $M \in \{2,4,8,16\}$, and each color corresponds to a different training methodology.
    Coverage is estimated through $N_\text{trials}=300$ trials, the solid lines are means over $N_\text{trainings}=10$ trainings, and the shaded regions correspond to the standard error over the trainings.
    The dashed line corresponds to ideal coverage, $c(z) = \bar{c}(z)$; the region above the dashed line corresponds to overcoverage, and the region below the dashed line corresponds to undercoverage.
    In each quadrant, the upper panel shows the coverage $c(z)$ and the lower panel shows the residual coverage $c(z) - \bar{c}(z)$.}
    \label{fig:training_protocols_gaussian}
\end{figure*}

To evaluate the behavior of $w_i f_i$ ensembles, we perform empirical tests of their coverage properties.
We emphasize that these (expensive) tests are only to validate the procedure, and they would not be needed to deploy $w_i f_i$ ensembles in practice.
Since the coverage depends on the flexibility of the model, we sweep over the number of trained networks $M \in \{2,4,8,16\}$.

For each of the protocols proposed in \Sec{subsec:nn_training} (Partition and Bootstrap) and each value of $M$, we train the networks $f_i$ $N_\text{trainings}$ times starting from different random initializations.
Then, for each one of these trainings, we perform $N_\text{trials}$ trials to fit the weights $w_i$ and extract confidence intervals on both $\log r$ and $\kappa$, where each trial consists of the following steps:
\begin{enumerate}
    \item Generate a dataset of size $N = \text{25,000}$.%
    \item Calculate $\hat{w}_i$ and $C_{ij}$ according to \Eq{eq:what} and \Eq{eq:sandwich}. We use the \texttt{pytorch-minimize} package \cite{torchmin} to perform full-batch optimization with Newton's method over the $w_i$.
    \item Sample one $x$ at which to evaluate $\log r$; we choose either $n$ or $d$ with probability $0.5$ and then sample $x$ from the chosen distribution.
    \item Calculate and save the true $\log r(x) = 2 \mu x$, the estimated $\log \hat{r}(x) = \hat{w}_i f_i(x)$, and the estimated variance $\sigma^2(x)$ of $\log \hat{r}(x)$ given by \Eq{eq:dre_unc}. The estimate and the variance can be used to obtain a $z \sigma$ confidence interval $\left[\hat{w}_i f_i(x) - z \sqrt{\sigma^2(x)}, \hat{w_i}f_i(x) + z \sqrt{\sigma^2(x)}\right]$.
    \item For each value of $\kappa$ in $\{0.01, 0.02, 0.05, 0.1, 0.2, 0.5\}$, generate a mixture dataset of size $N = \text{25,000}$ by sampling each event from $n$ with probability $\kappa$ and from $d$ with probability $(1-\kappa)$. 
    \item Compute $\hat{\kappa}$ from \Eq{eq:mf_pseudomle} and $\sigma^2_\text{GS}$ from \Eq{eq:GS}, using \Eq{eq:mix_frac_derivs} to evaluate the latter. We use the \texttt{minimize\_scalar} routine in SciPy~\cite{2020SciPy-NMeth} to perform the minimization and find $\hat{\kappa}$. 
    \item Construct $z \sigma$ confidence intervals as $\left[\hat{\kappa} - z \sqrt{\sigma_\text{GS}^2}, \hat{\kappa} + z \sqrt{\sigma_\text{GS}^2}\right]$.
\end{enumerate}

In the asymptotic limit, assuming the model is well-specified, the predictions for $\log \hat{r}$ ($\hat{\kappa}$) should be normally distributed around the true $\log r$ ($\kappa$) with variance equal to $\sigma^2$ ($\sigma^2_\text{GS}$).
For each training, we can estimate the coverage $c$ of the $z \sigma$ confidence intervals on a quantity $q$ with estimator $\hat{q}$ and uncertainty $\sigma^2_q$ as a function of $z$ as
\begin{equation}
\label{eq:coverage_estimate}
    c(z) = \frac{1}{N_\text{trials}} \sum_{i=1}^{N_\text{trials}} \Theta\left(z-\frac{\left|\hat{q} - q\right|}{\sqrt{\sigma^2_\text{q}(x_i)}} \right),
\end{equation}
where $\Theta$ is the Heaviside step function, defined to be $1$ when its argument is positive and $0$ otherwise.
Said in words, \Eq{eq:coverage_estimate} is the empirical fraction of outcomes observed in $N_\text{trials}$ that lie within the desired confidence interval.
The nominal coverage $\bar{c}(z)$ then satisfies $\bar{c}(1) \approx 0.68$, $\bar{c}(2) \approx 0.95$, and so forth; it is calculated as $\bar{c}(z) = P_\text{norm}(z) - P_\text{norm}(-z)$, where $P_\text{norm}$ is the cumulative distribution function of a one-dimensional standard normal distribution. 
Model misspecification and deviations from the asymptotic limit could cause departures from these nominal coverages. 
Comparing $c(z)$ to $\bar{c}(z)$ then provides a direct check of the coverage properties of confidence intervals.
To smooth out fluctuations over training instances, we average $c(z)$ over the $N_\text{trainings}$, and we also report the variances in the coverage over these instances.
For these experiments, we take $N_\text{trainings} = 10$ and $N_\text{trials} = 300$.

\subsection{DRE Results}
\label{subsec:gaussian_llr_results}

We start with the coverage properties on the density ratio itself.
In \Fig{fig:training_protocols_gaussian}, we plot the coverage $c(z)$ of $z \sigma$ confidence intervals on $\log r$.
The results for the Partition protocol are shown in blue and Bootstrap are in orange.
The nominal coverage $\bar{c}(z)$ is shown by the black dashed line.
Each quadrant of the plot shows a different value of $M$ (i.e.\ a different size ensemble).

The first phenomenon that we can see in  \Fig{fig:training_protocols_gaussian} is that the coverage improves as $M$ grows.
For small values of $M$, the confidence intervals tend to undercover, but as $M$ grows they saturate the correct coverage properties.
This is consistent with the interpretation that failure to achieve the nominal coverage comes from model misspecification; as $M$ grows, linear combinations of the $f_i$ become increasingly expressive, and model misspecification necessarily decreases.
We note that it is not always the case that coverage improves with increasing $M$; when $M$ becomes very large, asymptotic formulae derived with fixed $M$ and $N$ going to infinity are no longer applicable.
We show an example of this phenomenon, and describe how to correct for it with moderate values of $M$, in \App{app:bias}.

We can also observe in \Fig{fig:training_protocols_gaussian} that the differences in coverage between the training protocols are largely insignificant.
For $M=2$, $4$, and $16$, the difference between Bootstrap and Partition is so small that we cannot sharply distinguish their performance over $N_\text{trainings}=10$ trainings.
For $M=8$, Partition does clearly outperform Bootstrap, although in absolute terms the mean difference in coverage is never large; it is always less than $0.05$.
We also find that once the coverage has converged, i.e.\ in the $M=16$ case, the average sizes of the resultant confidence intervals are essentially identical for both methods.

\begin{figure}
    \centering
    \includegraphics[width=1.0\linewidth]{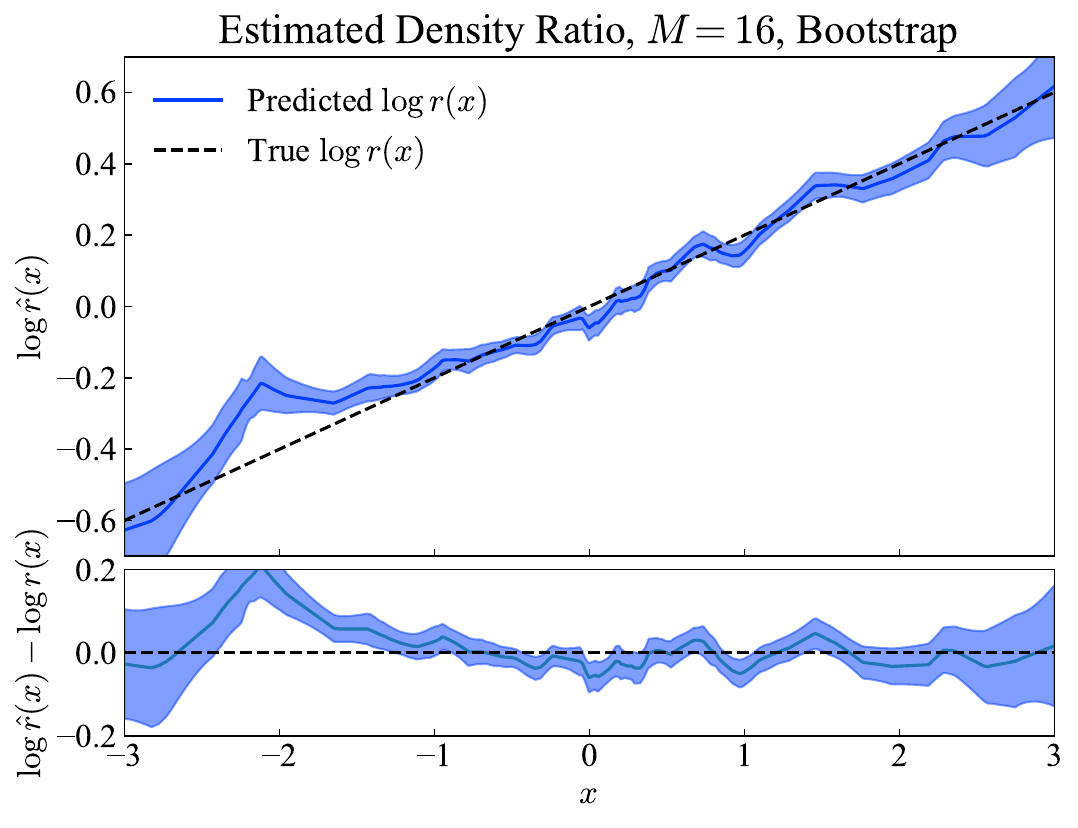}
    \caption{
    Upper panel: analytic (black dashed) and estimated (blue solid) log likelihood ratio $\log r$ as a function of $x$. The blue shaded region shows the $1 \sigma$ confidence interval for $\log r$ as a function of $x$.  Note that the confidence intervals are correlated between $x$ values.  Lower panel: the residual between the estimated and true log likelihood ratios.}
    \label{fig:llr}
\end{figure}

For concreteness, we explicitly show one estimator for $\log r(x)$ and its uncertainties in \Fig{fig:llr} with $M=16$, trained using the Bootstrap protocol.
A word of caution:  $\log \hat{r}(x) = \hat{w}_i f_i(x)$ and $\log \hat{r}(x') = \hat{w}_i f_i(x')$ are correlated, with covariance equal to $f_i(x) C_{ij} f_j(x')$.
This means that we do not necessarily expect $68 \%$ of the blue shaded region in \Fig{fig:llr} to cover the true $\log r$ for any particular instantiation of $\log \hat{r}$, as coverage for one value of $x$ is correlated with coverage for other values.
Rather, the relevant coverage property is that for each fixed $x$, $68 \%$ of intervals obtained in this way should contain the true value of $\log r$ for that $x$; this cannot be read off of \Fig{fig:llr}.

\begin{figure*}
    \centering
    \includegraphics[width=0.9\linewidth]{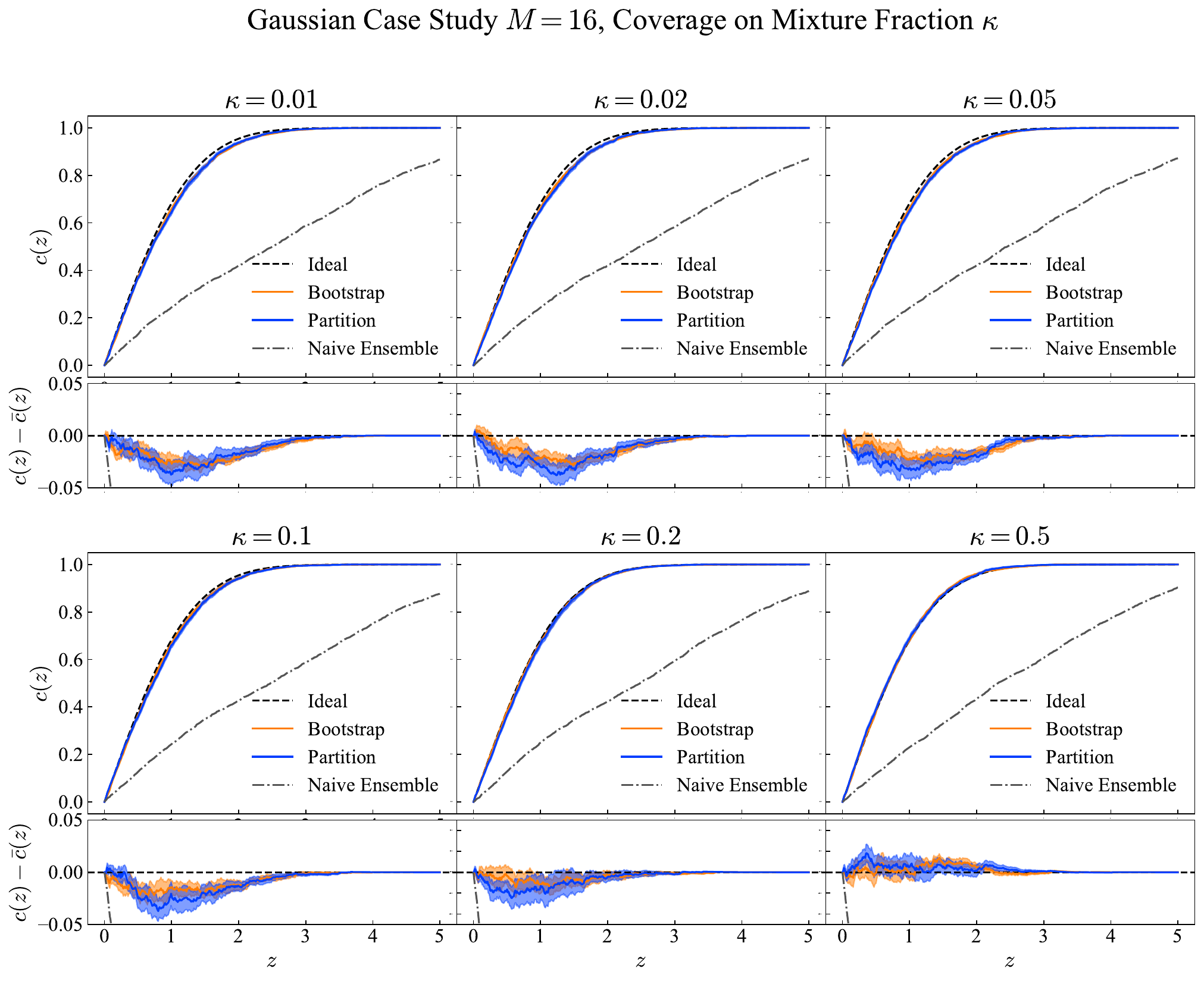}
    \caption{
    The same as \Fig{fig:training_protocols_gaussian}, but now assessing coverage on the mixture fraction $\kappa$ in the Gaussian case study with fixed $M = 16$.  The six displays correspond to $\kappa \in \{0.01, 0.02, 0.05, 0.1, 0.2, 0.5\}$.
    Also shown in the dot-dashed line is the Naive Ensemble protocol with fixed $w_i = 1/M$ and no quantification of DRE uncertainties.}
    \label{fig:f01_gaussian}
\end{figure*}

\subsection{Mixture Fraction Results}
\label{subsec:gaussian_mf_results}

We now move to the mixture fraction task, where the goal is to extract the value of $\kappa$.
In \Fig{fig:f01_gaussian}, we show the coverage results for all training methods with fixed $M = 16$, for each value of $\kappa \in \{0.01, 0.02, 0.05, 0.1, 0.2, 0.5\}$.
As with \Fig{fig:training_protocols_gaussian}, the solid lines and shaded regions show the coverage (or coverage residual) of each of the training methods.
The dot-dashed line labeled ``Naive Ensemble'', described in \Sec{subsec:nn_training}, corresponds to training an ensemble as in the Bootstrap protocol, but using uniform $1/M$ ensemble weights instead of fitting the $w_i$.
The uncertainties in this naive case are taken to be $\sigma^2_\text{MLE}$, so uncertainty on the density ratio itself is not assessed.

We can see that both the Partition and Bootstrap protocols achieve close to nominal coverage for all examined values of $\kappa$.
It is visible in the residuals that the coverage improves marginally as $\kappa$ becomes closer to its midpoint, and at $\kappa=0.5$ the curves are essentially indistinguishable from the ideal coverage curve.
Both $w_i f_i$ ensemble protocols dramatically outperform Naive Ensembles, which undercover on average for all examined values of $\kappa$.
We conclude that $w_i f_i$ ensembles yield reliable confidence intervals, at least in the context of a one-dimensional toy problem where the density ratio is known.

\section{Quark/Gluon Discrimination}
\label{sec:qg}

To evaluate the performance of $w_i f_i$ ensembles on a nontrivial task, we apply them to a paradigmatic task in QCD and jet substructure.
In this section, we use $w_i f_i$ ensembles to estimate the likelihood ratio of quark and gluon jets, as well as to estimate the quark fraction $\kappa$ in a mixed sample of quark and gluon jets.
In \Sec{subsec:qg_methods}, we explain the physics context, the dataset, and our methods in detail.
We present the results of this case study in \Sec{subsec:qg_results}, and we visualize the inferred likelihood ratio by reweighting a jet substructure observable in \Sec{sec:qg_dre_visualization}.

\begin{figure*}
    \centering    \includegraphics[width=0.9\linewidth]{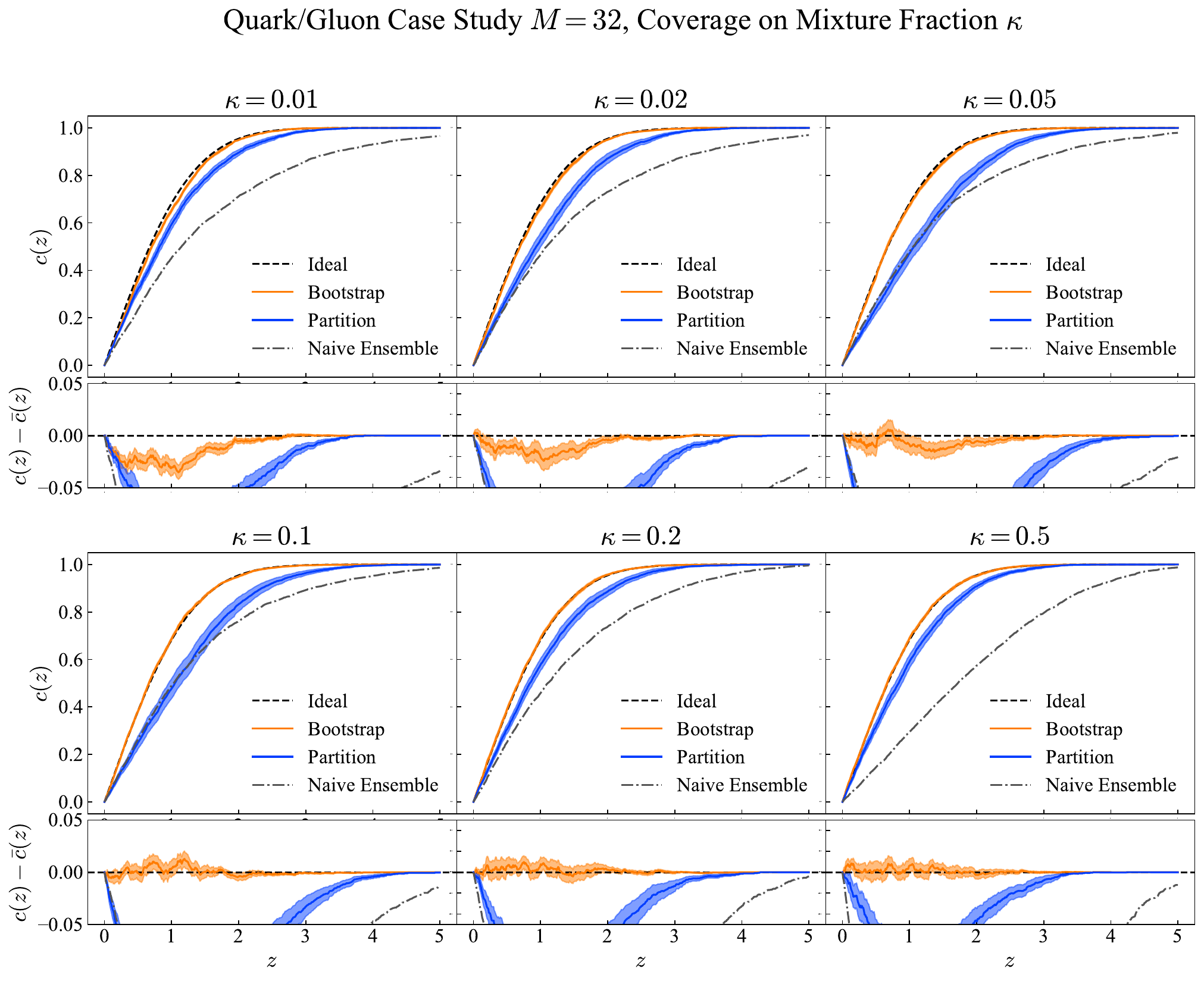}
    \caption{The same as \Fig{fig:f01_gaussian}, but now assessing coverage for the mixture fraction $\kappa$ in the quark/gluon case study with $M = 32$.}
    \label{fig:qg_results}
\end{figure*}

\subsection{Methodology}
\label{subsec:qg_methods}

The goal is to estimate the likelihood ratio of quark and gluon jets, which by the Neyman-Pearson lemma \cite{Neyman:1933wgr} is monotonically related to the optimal classifier for quark/gluon discrimination.
See \Reff{Cranmer:2015bka} for more on this equivalence, \Reffs{Jones:1988ay,Fodor:1989ir,Lonnblad:1990bi,Lonnblad:1990qp,Csabai:1990tg,Jones:1990rz,Pumplin:1991kc,OPAL:1993uun, Gallicchio:2012ez, Gras:2017jty} for a selection of work on quark/gluon discrimination, and \Reffs{Larkoski:2017jix, Kogler:2018hem} for reviews of jet substructure in general.
We focus only on information that is infrared and collinear (IRC) safe, so strictly speaking the extracted likelihood ratio is only monotonically related to the optimal IRC-safe classifier.

For this case study, we will not be able to directly evaluate coverage for the likelihood ratio, as the true likelihood ratio is not explicitly known.
We therefore only examine performance on the mixture fraction task, where $\kappa$ is now the quark fraction of a mixed sample of quark and gluon jets.
We use the quark/gluon dataset \cite{Zenodo:EnergyFlow:Pythia8QGs} included with the \texttt{energyflow} package \cite{Komiske:2018cqr}, which was generated using Pythia 8.226 \cite{Bierlich:2022pfr}.
This dataset contains $2$ million total anti-$k_t$ \cite{Cacciari:2008gp} jets clustered with jet radius $R = 0.4$ using \textsc{FastJet} \text{3.3.0}~\cite{Cacciari:2011ma}, restricted to transverse momenta with $p_T \in \{500,550\}\, \text{GeV}$.

For the  $f_i$, we use Energy Flow Networks (EFNs) \cite{Komiske:2018cqr}, which are naturally suited to the point cloud format of collider events and which respect IRC safety.
We note that IRC safety is motivated by, but insufficient for, perturbative tractability (see \Reff{Bright-Thonney:2023gdl} for more details in the context of EFNs).
As before, we implement EFNs in PyTorch \cite{paszke2019pytorchimperativestylehighperformance}.
An EFN has a per-jet $\Phi$ network and a per-particle $F$ network, and we take $\Phi$ to have two hidden layers and $F$ to have three, with all hidden layers taken to have width $32$ for simplicity.
We use Leaky ReLU activation functions with leakiness $0.2$.
We also take the latent dimensionality of $\Phi$ to be $32$, since this is where \Reff{Komiske:2018cqr} begins to observe performance saturation for the quark/gluon task.

\begin{figure*}
   \centering    \includegraphics[width=0.9\linewidth]{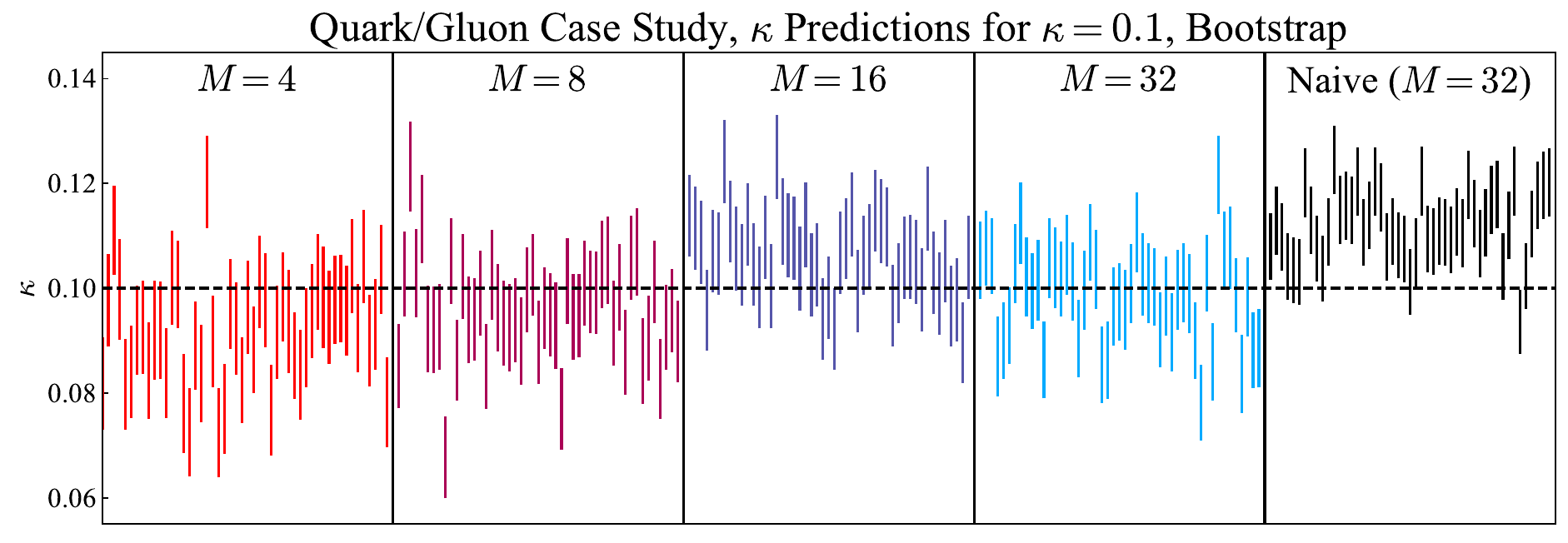}
    \caption{For the quark/gluon case study, results from $50$ mixture fraction predictions and intervals each for $M \in \{4, 8, 16, 32\}$.
    The rightmost panel corresponds to Naive Ensemble predictions, and the remaining panels are $w_i f_i$ ensembles trained with the Bootstrap protocol. 
    For each value of $M$, all of the predictions come from the same initial training but different mixture data sets for inference.
    The horizontal axis is arbitrary; each value on this axis corresponds to one prediction of (and interval on) $\kappa$. The vertical axis shows the values of $\kappa$ lying within each interval. The dashed horizontal line corresponds to the true value $\kappa=0.1$.}
    \label{fig:example_intervals}    
\end{figure*}

We sweep over the number of networks $M \in \{4,8,16,32\}$, and again consider the Partition and Bootstrap protocols, along with the Naive Ensemble protocol as a baseline.
As before, we perform $N_\text{trainings} = 10$ rounds of training and $N_\text{trials} = 300$ rounds of fitting the $\hat{w}_i$ to training data and estimating mixture fractions $\kappa$ at the same value as in \Sec{sec:gaussian}: $\kappa \in \{ 0.01, 0.02, 0.05, 0.1, 0.2, 0.5\}$.

Since this case study involves a fixed data set, we cannot generate new events on the fly as in the Gaussian example.
To ensure that the data sets used for training and inference are totally disjoint, for each protocol and for each training, we randomly divide the total dataset in half to obtain two disjoint pools of events.
We sample $\text{20,000}$ jets from the first pool (so  $N_n \approx N_d \approx \text{10,000}$) to obtain the training set, which we use to train the $f_i$ (again using early stopping with a patience of $10$ using a disjoint validation set of the same size, also drawn from the first pool).
For each trial, we sample a new dataset (of the same size) again from the first pool to fit the $\hat{w}$.
Then, for each value of $\kappa$, we draw a mixture dataset of $\text{10,000}$ events from the second pool to use for inference.

We use analytic formulae for the gradient and Hessian of \Eq{eq:sample_symMLC} to fit the $\hat{w}$, using the \texttt{minimize} routine included in SciPy~\cite{2020SciPy-NMeth} to perform Newton's method.
We again use the \texttt{minimize\_scalar} routine to calculate $\hat{\kappa}$ and its confidence interval.

\subsection{Mixture Fraction Results}
\label{subsec:qg_results}

In \Fig{fig:qg_results}, we show the coverages for each value of the mixture fraction $\kappa$ with $M=32$, which is the value of $M$ at which performance saturates.
In contrast with the behavior observed in \Sec{sec:gaussian}, there is a stark dependence on training methodology here: Bootstrap clearly outperforms Partition for all values of $\kappa$.
This makes some intuitive sense; in this higher-dimensional setting, it seems plausible that linear combinations of very poorly trained functions (as one obtains after partitioning the data $32$ times, for example) are insufficiently expressive to model the data generating process, while this behavior is masked in a low-dimensional toy problem.
These results suggest that well-trained networks may become increasingly important to reach nominal coverage as the problem becomes increasingly high-dimensional.
We have also checked that the coverage is not meaningfully different from the $\kappa=0.01$ case for smaller values of $\kappa$.

For concreteness, we show $50$ example predictions and intervals for $\kappa = 0.1$, $M \in \{4,8,16,32\}$ using Bootstrap in \Fig{fig:example_intervals}.
These results demonstrate that even the smallest $w_i f_i$ ensembles we consider produce qualitatively sensible intervals (i.e.\ they are in the vicinity of the correct value of $\kappa$ and they cover a significantly nonzero proportion of the time), but only sufficiently large ensembles ($M = 32$ for this dataset) have sufficient expressivity to produce quantitatively correct coverage properties.
We also show predictions from a Naive Ensemble with $M=32$; in contrast with the $w_i f_i$ ensembles, these intervals clearly undercover and show a substantial bias.
The magnitude and sign of this bias varies from training to training.
The training depicted in \Fig{fig:example_intervals} is representative: the magnitude of the bias in this training is close to the average bias we observe over the $10$ trainings that we performed.
The presence of this bias for the Naive Ensemble and its absence for the $w_i f_i$ ensembles shows that $w_i f_i$ ensembles not only allow us to characterize uncertainties, but also yield a better point estimate of the density ratio than the Naive Ensemble.
Indeed, we observe that for fixed $f_i(x)$, the average values over $N_\text{trials}$ trials of the $\hat{w}_i$ vary widely, and are not generically well-approximated as $1/M$.
Specifically, we observe that the average $\hat{w}_i$ still approximately sum to $1$, but that they are sometimes negative and span a range of magnitudes from $O(10^{-2})$ to $O(1)$.

\subsection{DRE Visualization}
\label{sec:qg_dre_visualization}

Since the quark/gluon likelihood ratio is jet-valued and therefore high dimensional, we cannot visualize it directly.
That said, we can provide a visualization of the learned likelihood ratio that is somewhat in the spirit of \Fig{fig:llr}. 
We consider the distribution of a particular observable, the jet angularity~\cite{Berger:2003iw,Almeida:2008yp,Ellis:2010rwa,Larkoski:2014pca}
\begin{equation}
    \lambda_{\beta} = \sum_i z_i \theta_i^\beta,
\end{equation}
where the sum is over jet constituents, $z_i$ is the transverse momentum fraction of constituent $i$ (i.e. $p_{Ti} / \sum_j p_{Tj}$), $\theta_i = \sqrt{y_i^2 + \phi_i^2}$ is its angular distance from the jet axis (which we take here to be the $p_T$ weighted average position of the jet), $y_i$ is the rapidity, and $\phi_i$ is the azimuthal coordinate.
We specialize to the $\beta=2$ angularity, which is related to the jet mass.
\begin{figure}
    \centering
    \includegraphics[width=1.0\linewidth]{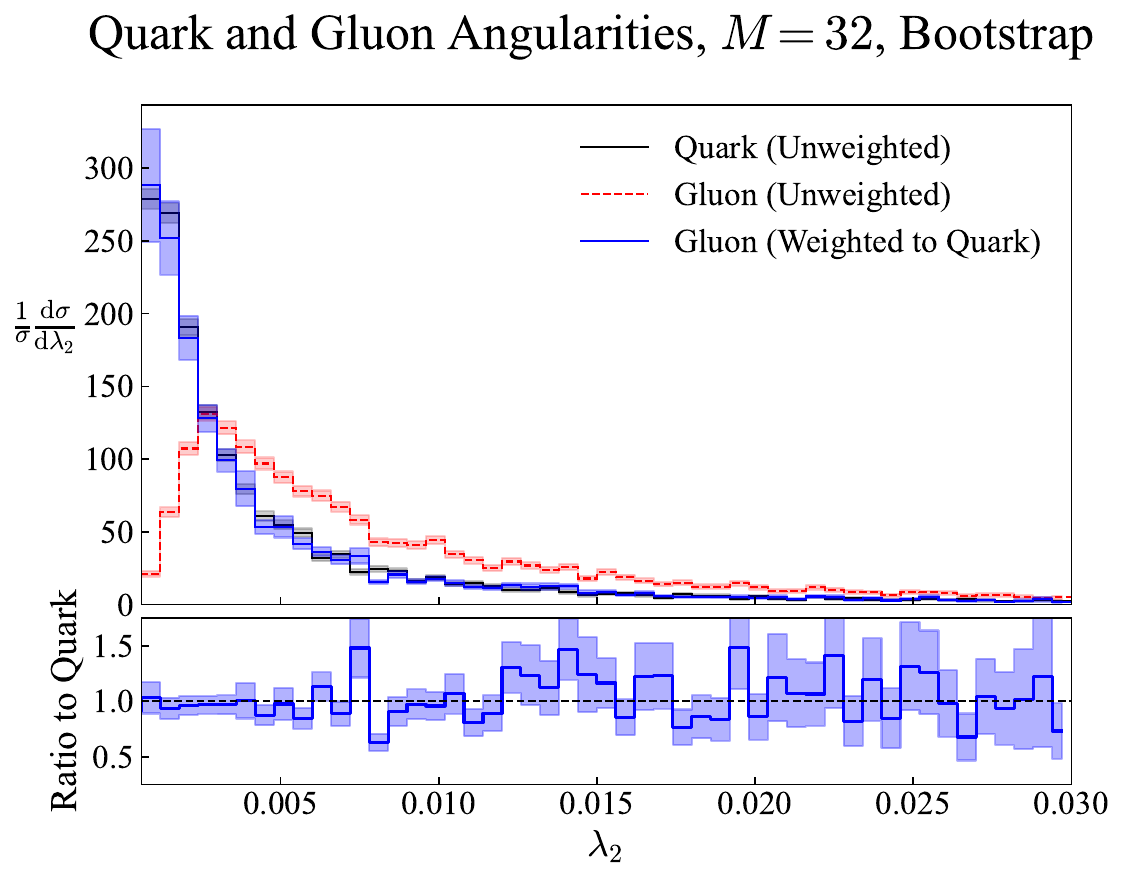}
    \caption{Distributions of the $\beta=2$ angularities for quark and gluon jets in our simulated dataset. The top panel shows the distribution of the angularities themselves for quark (black) and gluon (red) jets with unit weights, as well as the distribution for gluons jet reweighted using the learned quark/gluon density ratio (blue). The displayed uncertainties correspond both to Poisson uncertainties (for all distributions) and to the uncertainties in the weights (for the reweighted distribution).}
    \label{fig:angularity}
\end{figure}

We show the distribution of this angularity for quark and gluon jets in \Fig{fig:angularity}.
In the top panel, the red and black curves correspond to the results for gluon and quark jets respectively, and the blue curve shows gluon jets reweighted by the learned quark/gluon likelihood ratio, using an $M=32$ Bootstrap $w_i f_i$ ensemble.
In the bottom panel, the blue curve shows the ratio of this reweighted distribution to the quark distribution.

The uncertainty bands are $1 \sigma$ confidence intervals including both the Poisson statistical uncertainty as well as the uncertainty in the learned density ratio from the $w_i f_i$ ensemble.
We calculate the former as the square root of the sum of squared weights in each bin, and the latter by sampling $3000$ values of the $w_i$ from a Gaussian centered at $\hat{w}_i$ with covariance $C_{ij}$, and calculating the standard deviation over samples of the sum of weights in each bin.\footnote{These calculations are performed at the level of counts, since it is counts that are Poisson distributed; the unit conversion to a probability density is performed at the end.}
It can be shown that these uncertainties add in quadrature up to terms subleading either in $C$ or the number of counts in each bin.

We observe good agreement between the quark curve and the gluon-reweighted-to-quark curve.
Since the reweighting introduces correlations between the bins, quantitatively checking the coverage of the intervals would require reweighting an ensemble of datasets; one cannot simply count the proportion of bins with intervals that cover for any given dataset.
However, the uncertainties are qualitatively reasonable, covering in most but not all bins.

\section{Conclusions}
\label{sec:conclusions}

In this paper, we introduced $w_i f_i$ ensembles, a new method for estimating frequentist confidence intervals both for density ratio estimation tasks and for downstream parameter estimation.
We derived asymptotic coverage guarantees for these confidence intervals under the assumption that the $w_i f_i$ model is well-specified.
We then tested these intervals empirically, both in a Gaussian toy problem and in a nontrivial jet substructure application, and we found that the intervals can successfully achieve the desired coverage properties when the ensemble is sufficiently large.
For completeness, we also derive the leading order bias of the Gong-Samaniego estimator in \App{app:bias}, which is relevant for two-step inference procedures more broadly.

We tested two different training protocols for $w_i f_i$ ensembles and found that the Bootstrap approach gave the best overall performance, but there are many more possibilities to be explored.
We did not consider ensembles of different types of models, or of models trained with different hyperparameters; it is plausible that these directions could yield more diverse ensembles and achieve satisfactory coverage with smaller ensemble sizes $M$.
Moreover, many methods exist in the literature to obtain ensembles of models without the naive $O(M)$ computational scaling, like \Reffs{huang:2017, garipov:2018, wen:2020}.
It would be interesting to examine $w_i f_i$ ensembles trained with efficient protocols such as these.

Since the $f_i(x)$ must be fixed for the inference on the $w_i$ to be well defined, we have used separate datasets $D_\text{train}$ and $D_\text{fit}$ to train the $f_i(x)$ and to fit the $w_i$.
In order to achieve coverage and maximize power (i.e.\ minimize the size of intervals) given a fixed budget of data, there is then an interesting tradeoff to make between allocating data to $D_\text{train}$ and $D_\text{fit}$.
In particular, the quantified uncertainties are only directly sensitive to the size of $D_\text{fit}$, but an insufficiently large $D_\text{train}$ may result in poor basis functions $f_i(x)$ and cause model misspecification.
Using the same dataset to train the $f_i(x)$ and to fit the $w_i$ would remove this tradeoff and allow one to use all of the data both for training the $f_i(x)$ and for fitting the $w_i$, but in principle correlations between the $w_i$ and the $f_i(x)$ must then be taken into account, and without accounting for these correlations we are unable to rigorously guarantee coverage when reusing the training dataset to fit the $w_i$.
In the cases that we have checked, however, neglecting these correlations does not seem to affect the coverage properties of our estimators.
It would be interesting to investigate these correlations further, and particularly to determine how to include their contributions to uncertainties and when it is safe to neglect them.

It would also be interesting to investigate how $w_i f_i$ ensembles can slot into existing workflows.
In principle, if one already has a well-trained estimator of the density ratio, one can simply use this model as one of the $f_i$ and train the remaining networks using some other, less computationally demanding protocol (or using other estimators that one may already have at hand).
In the limit where training this powerful estimator takes a very large amount of computation relative to evaluating it, this could be used to obtain a notion of uncertainties with a comparatively small amount of additional compute.
If one already has an ensemble of estimators, converting this preexisting ensemble to a $w_i f_i$ ensemble only requires the optimization in \Eq{eq:what} to obtain the $\hat{w}_i$.
Additionally, while the formulae for calculating uncertainties that we have presented throughout are only applicable in the asymptotic regime, the $w_i f_i$ ensemble ansatz could be combined with existing methods (like the Neyman construction) to yield small-sample uncertainties on density ratios.

While we proposed $w_i f_i$ ensembles in the context of extracting reliable confidence intervals, they may also be useful even if one is not interested in obtaining uncertainties.
In particular, it seems plausible that a $w_i f_i$ ensemble may yield a better point estimate of the density ratio than a traditional ensemble; fitting the $\hat{w}_i$ provides a data-driven way to lend more credence to the members of the ensemble that are more helpful and ignore those that are not.
This could be a useful supplement to existing techniques using ensembles to stabilize DRE, like those explored in \Reff{Acosta:2025lsu}.

We observe hints of improved stabilization in our experiments.
In the context of the quark fraction task, we found that Bootstrap not only yields confidence intervals with much better coverage properties than the (identically trained) Naive Ensemble, but it also provides point estimates for $\hat{\kappa}$ with a substantially smaller mean squared error.
In other words, the performance difference between Bootstrap and Naive Ensemble is not only due to the need to account for uncertainties in the density ratio to achieve reliable uncertainties on the parameter $\kappa$, but also because Bootstrap does a substantially better job at estimating the density ratio (and therefore $\kappa$) in the first place.
It would be interesting to explore this further, to determine how general this improvement might be.
Example implementations of $w_i f_i$ ensembles, and in particular the code used to generate the figures, can be found at Ref.~\cite{benevedes:wifiGithub}.
\acknowledgments

We would like to thank Brian Nord for inspiring conversations about uncertainty quantification,  Aishik Ghosh for discussion about ensemble methods, and Prasanth Shyamsundar for discussion about correlations between the $f_i$ and the $w_i$.
We would also like to thank Rikab Gambhir and Benjamin Nachman for useful discussions and comments.
SB and JT are supported by the National Science Foundation (NSF) under Cooperative Agreement PHY-2019786 (The NSF AI Institute for Artificial Intelligence and Fundamental Interactions, \url{http://iaifi.org/}), and by the U.S. Department of Energy (DOE) Office of High Energy Physics under grant number DE-SC0012567. JT is additionally supported by the Simons Foundation through Investigator grant 929241, and in part by grant NSF PHY-2309135 to the Kavli Institute for Theoretical Physics (KITP). The computations in this paper were run on the FASRC Cannon cluster supported by the FAS Division of Science Research Computing Group at Harvard University.

\appendix

\section{Properties of the \texorpdfstring{$\hat{w}$}{w hat} Estimator}
\label{app:what}

In this Appendix, we expound on crucial properties of the estimated weights $\hat{w}_i$, including their normality and asymptotic covariance matrix, as needed for the discussion in \Sec{subsec:est_cij}.
We assume throughout that the model is well-specified, meaning that there is some true value $w_i^*$ of the $w_i$ for which $w_i^* f_i = \log r$, for $r$ the density ratio of interest.

Recall that the $\hat{w}_i$ are defined as the minimizers of the $\mathcal{L}_{\text{samp}}$ loss in \Eq{eq:sample_symMLC}.
Because the loss is defined through a sample average, this estimator is then an $M$-estimator in the language of~\Reff{huber64}.
It is consistent (i.e.\ asymptotically unbiased), since in the large $N$ limit $\mathcal{L}_\text{samp}$ converges to $\mathcal{L}_\text{symMLC}$ in \Eq{eq:symMLC}, which is minimized by the $w^*_i$.

Defining the gradient with respect to the weights as $u_i(w) \equiv \partial_{w_i}\mathcal{L}_\text{samp}$, the minimum $\hat{w}$ satisfies
\begin{equation}
    u_i(\hat{w}) = 0.
\end{equation}
Performing a Taylor expansion of this expression around the true $w^*$, we have
\begin{align}
    0 = u_i(\hat{w}) = u_i(w^*) &+ (w^*_j - \hat{w}_j)u_{ij}(w^*) \nonumber\\
    &+ O\big((w^* - \hat{w})^2\big),
\end{align}
where we have defined $u_{ij} \equiv \partial_{w_j} u_i$.
Asymptotically we are justified in neglecting the quadratic and higher order terms due to the consistency of $\hat{w}$, so we can rearrange this expression as
\begin{equation}
\label{eq:pre_normality_covariance}
    \sqrt{N} (w^*_j - \hat{w}_j) = u^{ji}(w^*) \big(\sqrt{N}u_i(w^*)\big),
\end{equation}
where $u^{ji}$ is the matrix inverse of $u_{ji}$.
The factors of $N$ (i.e.\ the characteristic data size) have been inserted for later convenience, and we have restricted for simplicity to the case of balanced samples $N_n = N_d \equiv N$.

The central limit theorem ensures that $ \sqrt{N} u_i(w^*)$ is normally distributed; it is easy to check that its mean is zero for our estimator, and we denote its covariance matrix as $\tilde{U}_{ik}$.
This covariance matrix does not scale with $N$.
On the other hand, due to the law of large numbers applied to $u_{ji}(w^*)$, $u^{ji}(w^*)$ asymptotes to its expectation value $V^{ji}(w^*)$, where $V^{ij} \equiv V_{ij}^{-1}$.
Slutsky's lemma~\cite{SlutsukyberSA}, which allows us to apply these limiting arguments independently to each factor in a product, then tells us that the left-hand side of \Eq{eq:pre_normality_covariance} is normally distributed as:
\begin{equation}
    \sqrt{N} (w^*_j - \hat{w}_j) \sim \mathcal{N}(0, V^{ik}\tilde{U}_{kl}V^{lm}).
\end{equation}
Defining $U_{ij} = \tilde{U}_{ij}/N$ to match the notation in the main text, we have
\begin{equation}
    (w^*_j - \hat{w}_j) \sim \mathcal{N}(0, V^{ik}U_{kl}V^{lm}).
\end{equation}
Thus, the estimated weights are indeed normally distributed with covariance matrix given by \Eq{eq:sandwich}, which as mentioned in \Sec{subsec:est_cij} can itself be consistently estimated from the data.

\section{Pseudo-likelihood Ratio Test}
\label{app:plrt}

There are a variety of ways to construct confidence intervals in standard asymptotic likelihood theory.
In the main text, we focused for simplicity on the analog of Wald intervals~\cite{wald1943}, which are obtained in the one-dimensional case as ``central value plus or minus estimated standard deviation''.

In the collider physics community, however, it is common practice is to construct intervals using a likelihood ratio test.
Specifically, it can be shown that (twice) the difference between the log-likelihood evaluated at the maximum likelihood estimator and evaluated at the true value of the parameters follows a $\chi^2$ distribution, so this difference can be used as a test statistic to construct confidence intervals.
Intervals constructed with the likelihood ratio test enjoy several theoretical advantages, notably including parametrization invariance.

As such, in this Appendix, we review an analogous pseudo-likelihood ratio test appropriate for the construction of confidence intervals on a single parameter $\theta$.
This test, and the general test appropriate for inference over multiple parameters, was proposed in \Reff{plrt}.
In our empirical case studies, the intervals produced with the pseudo-likelihood ratio test were essentially indistinguishable from the naive intervals we presented in the main text, but we provide information on this alternative approach for completeness.

Consider the Taylor expansion around the estimated parameter $\hat{\theta}$ of the log-pseudo-likelihood evaluated at $\hat{w}$ and the ground truth value $\theta^*$:
\begin{align}
    \log L(\Phi | \hat{w}, \theta^*) \approx  &\log L(\Phi|\hat{w}, \hat{\theta}) \nonumber \\
    + \frac{1}{2}(\theta^* &- \hat{\theta})^2 \left (\frac{\partial^2}{\partial \theta^2} \log L(\Phi| \hat{w}, \theta) \right) \biggr \rvert_{\theta = \hat{\theta}},
\end{align}
where the linear term vanishes since $\hat{\theta}$ is at a maximum of the pseudo-likelihood, and the higher order terms can be neglected since they are subleading in the characteristic dataset size $N$.
The distribution of $\hat{\theta} - \theta^*$ is asymptotically Gaussian with mean $0$ and variance $\sigma^2_\text{GS}$ given by \Eq{eq:GS}, and the second derivative term can be replaced at this order by minus the inverse of $\sigma^{2}_\text{MLE}$.
This means that the test statistic $T$ defined as
\begin{equation}
    T \equiv\frac{2 \left(\log L(\Phi|\hat{w}, \hat{\theta}) - \log {L(\Phi|\hat{w}, \theta^*)}\right)}{1 + \sigma^2_\text{MLE}A_i C_{ij} A_j}
\end{equation}
is asymptotically $\chi^2$ distributed with one degree of freedom (with $A$ and $C$ defined as in \Secs{subsec:param_estimation}{subsec:est_cij}).
As expected, this reduces to the usual likelihood ratio test statistic in the limit where the uncertainty on the $w$ goes to zero.

\section{Leading Asymptotic Bias}
\label{app:bias}

The estimators for the weights $\hat{w}$ and fractions $\hat{\kappa}$ presented in \Sec{sec:method} are asymptotically unbiased in the sense that their bias vanishes at the leading order in the asymptotic expansion, i.e.\ it is $o(N^{-1/2})$, but it does not vanish at higher orders.
The leading contribution to the bias of these estimators is generically $O(N^{-1})$.

Naively, the squared bias should then be negligible compared to the variance for $N \gg 1$.
This is borne out in most of our experimentation, but we find that this expectation can break down as the number of members of the ensemble $M$ grows large.
This is because the asymptotic expansion assumes that all other quantities in the problem, like parameter dimensionality, are held fixed and finite as $N$ grows large, so that in this power counting $N \gg M^p$ for any fixed $p$.

When this assumption is only mildly violated, the asymptotic expansion does not totally break down, but higher order terms can become important.
In our experiments, this manifests as a bias in $\hat{\kappa}$, which causes coverage to be nonmonotonic in $M$ for a fixed $N$ (i.e.\ as $M$ gets big, coverage first gets better, then worse).

In this Appendix, we quote the result for the leading order bias of an $M$-estimator, which immediately yields the bias on the $\hat{w}$.
We also derive the leading order bias in the Gong-Samaniego estimator for $\hat{\kappa}$.
We study an example where coverage of a $w_i f_i$ ensemble is adversely affected by the presence of a bias, that this bias is well-described by the leading order asymptotic bias, and that subtracting off an estimate of this bias largely restores coverage for $\hat{\kappa}$.

Formally, the next-to-leading order contribution to the variance is of the same order as the leading squared-bias, but we find that it is numerically less important for coverage in our experiments.
As such, we leave the calculation of this contribution to the variance (which is substantially more involved than the calculation of the bias) to future work.

The study of higher-order contributions to the distribution of asymptotic estimators has a long history.
The asymptotic bias of the MLE was first calculated in 1968 in \Reff{coxsnell}, and the next-to-leading contribution to the variance was first calculated in \Reff{mlevar}.
There has been a large amount of subsequent work on so-called high-dimensional statistics, where the dimensionality of the parameters is allowed to grow with $N$; see for example \Reff{Sur_2019}, which considers the context of high-dimensional logistic regression.

Within this large body of work, the leading order bias for an $M$-estimator has been studied in Refs.~\cite{RILSTONE1996369, econometrics4040048}.
This immediately allows us to estimate the bias of $\hat{w_i}$ as:
\begin{align}
    \langle \hat{w}^i - w^{*i} \rangle &= V^{ik} \bigg(V^{lm}\sum_{x} u_{s,kl}u_{s,m} \nonumber\\
    &+ \frac{1}{2}V^{ln}V^{mp} \Big(\sum_x u_{s,klm} \Big) \Big( \sum_x\ u_{s,n}u_{s,p} \Big) \bigg),
\end{align}
where the $s$ subscript indicates that the quantity is evaluated on one sample, not the whole dataset, and other subscripts indicate differentiation with respect to the $w_i$.
All quantities can be calculated with $w_i = \hat{w}_i$ at this order.
Empirically, we find that this correction is negligible throughout.
\begin{figure*}
    \centering
    \includegraphics[width=0.9\linewidth]{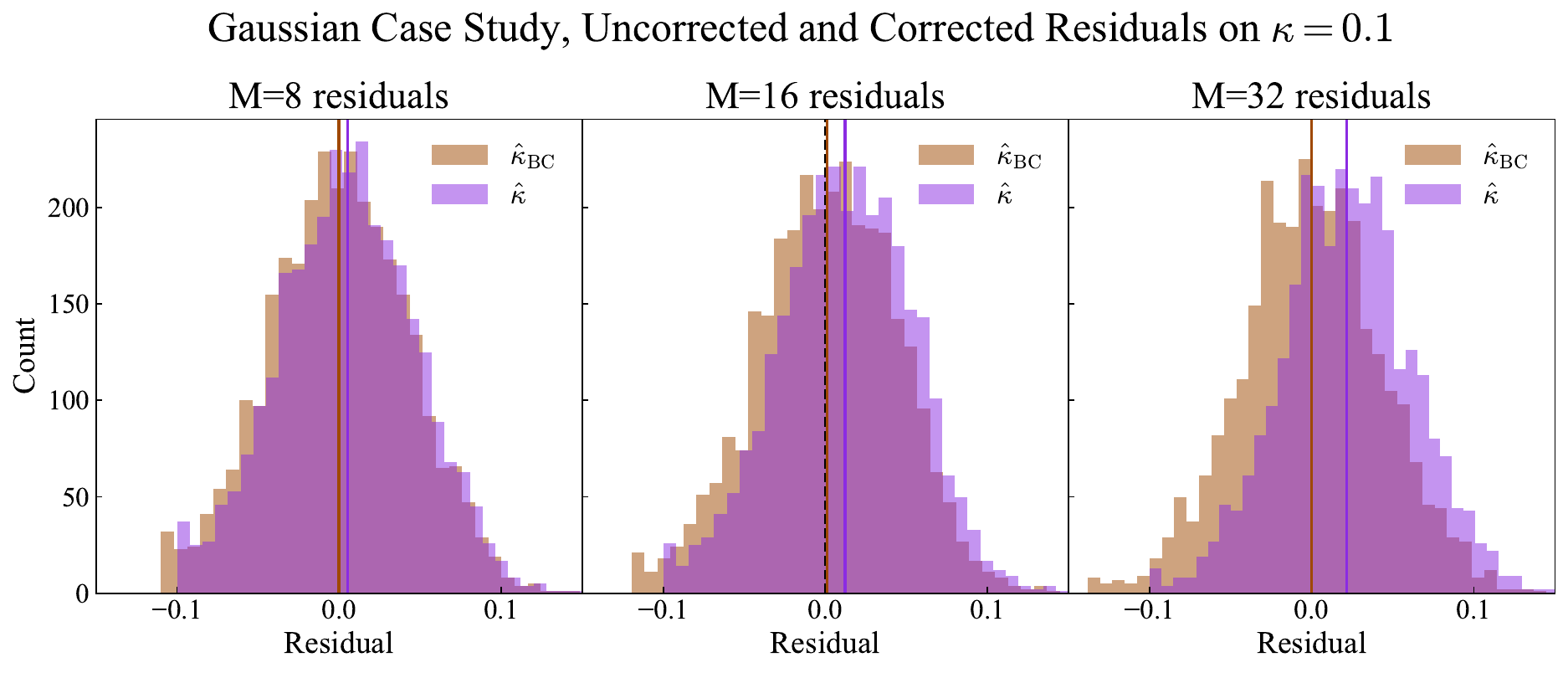}
    \caption{
    Residuals histogram for the mixture fraction task in the Gaussian toy problem with $\kappa=0.1$, using $M \in \{8, 16, 32\}$ networks in the \{left, middle, right\} panel. The networks are trained using the Bootstrap protocol with (brown) and without (purple) the bias correction. The dashed black vertical line is at $0$, the brown vertical line is the mean of the bias corrected predictions, and the purple vertical line is the mean of the uncorrected predictions; results are for $10$ trainings with $300$ trials each.}
    \label{fig:residuals_bias}
\end{figure*}
For the pseudo-likelihood estimator for a parameter $\theta$, where the likelihood is a function of parameters $w_i$ and the pseudo-likelihood is obtained with the plug-in estimators $\hat{w_i}$, we can Taylor expand the pseudo-score $u$ to find
\begin{align}
\label{eq:asymp_expansion}
    u(\hat{\theta}, \hat{w}) &= u + (\hat{\theta} - \theta^*) u' + (\hat{w}_i - w^*_i) u_i \nonumber\\
    &+ \frac{1}{2} (\hat{\theta} - \theta^*)^2 u'' + \frac{1}{2} (\hat{\theta} - \theta^*) (\hat{w}_i - w^*_i) u_i'\nonumber\\
    &+ \frac{1}{2}(\hat{w}_i - w^*_i)(\hat{w}_j - w^*_j) u_{ij} + O(N^{-1/2}),
\end{align}
where primes denote differentiation with respect to $\theta$, subscripts mean differentiation with respect to the $w_i$, and the functions on the right-hand side are all evaluated at the true parameter values.
By definition of the pseudo-MLE, the left-hand side vanishes, so at leading order we find
\begin{equation}
    \hat{\theta} - \theta^* = \frac{-u - (\hat{w}_i - w_i^*)u_i}{u'},
\end{equation}
and by the law of large numbers, at this order of approximation we can substitute the denominator with its expectation value.
We now plug this into \Eq{eq:asymp_expansion}, use that the left-hand side vanishes, and take an expectation value to obtain
\begin{align}
    \langle(\hat{\theta} - \theta^*)u'\rangle = -\frac{1}{2}&\bigg[\frac{\langle u^2 u'' \rangle}{\langle u' \rangle^2}  + \frac{C_{ij} \langle u_i u_j u'' \rangle}{\langle u' \rangle^2} \nonumber\\
    &- \frac{C_{ij} \langle u_i u'_j \rangle}{\langle u'\rangle} + C_{ij} \langle u_{ij} \rangle\bigg].
\end{align}
Furthermore, we have
\begin{align}
    \langle (\hat{\theta} - \theta^*) u'\rangle &= \langle\hat{\theta} - \theta^*\rangle \langle u' \rangle + \text{Cov}[\hat{\theta} - \theta^*, u'] \nonumber\\
    &=  \langle\hat{\theta} - \theta^*\rangle \langle u' \rangle - \text{Cov}\left[\frac{u}{\langle u' \rangle}, u'\right]\nonumber \\
    &=\langle\hat{\theta} - \theta^*\rangle \langle u' \rangle - \frac{\langle u u'\rangle}{\langle u' \rangle},
\end{align}
so
\begin{align}
    \langle \hat{\theta} - \theta^* \rangle = \frac{\langle u u' \rangle}{\langle u' \rangle^2} - \frac{1}{2 \langle u' \rangle}\bigg[\frac{\langle u^2 u'' \rangle}{\langle u' \rangle^2}  + \frac{C_{ij} \langle u_i u_j u'' \rangle}{\langle u' \rangle^2}\nonumber\\
    - \frac{C_{ij} \langle u_i u'_j \rangle}{\langle u'\rangle} + C_{ij} \langle u_{ij} \rangle\bigg].
\end{align}

\begin{figure*}
    \centering
    \includegraphics[width=0.9\linewidth]{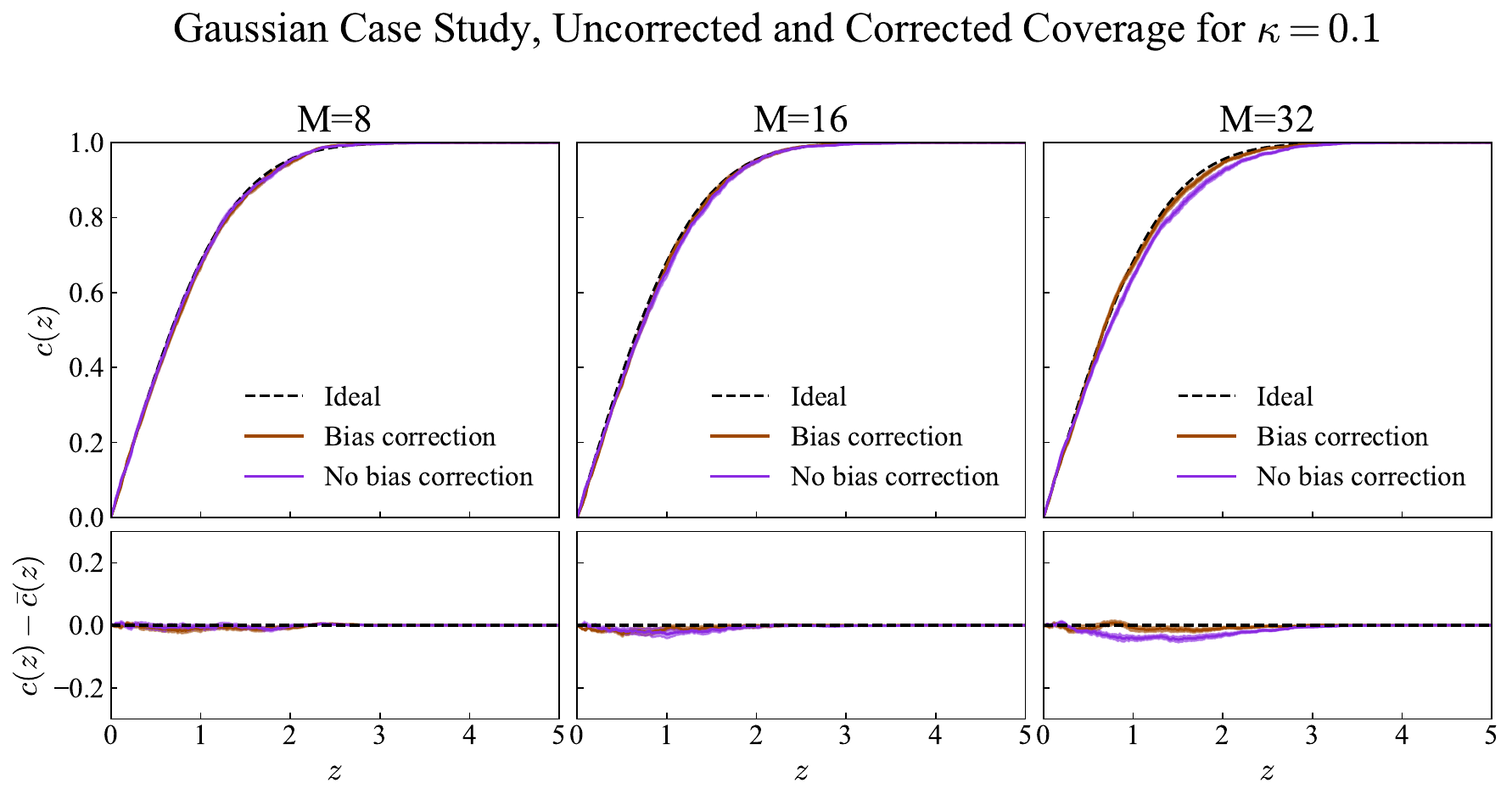}
    \caption{Coverage plot for the mixture fraction task in the Gaussian toy problem with $\kappa = 0.1$ using $M \in \{8, 16, 32\}$ networks trained with the Bootstrap protocol with (brown) and without (purple) the bias correction. Results are shown for $10$ trainings with $300$ trials each. The dashed black line in each panel shows the nominal coverage; the top panels show coverage, and the bottom panels show the difference between the coverage and the nominal coverage.}
    \label{fig:bias_correction}
\end{figure*}

We can now do some decompositions.
Let $a,b,c\ldots$ be sample indices, then for example we have
\begin{equation}
u = \frac{1}{N} \sum_a u_a,
\end{equation}
and likewise for all derivatives of $u$.
Therefore,
\begin{align}
    \langle u^2 u'' \rangle &= \left \langle \sum_{a,b,c} u_a u_b u''_c \right\rangle \nonumber \\
    &= N^2 \langle u_s^2 \rangle \langle u''_s \rangle^2 + O(N),
\end{align}
where the subscript $s$ means that the function is evaluated on one sample rather than on the entire dataset, and the $N^3$ term vanishes because $\langle u_s \rangle = 0$.
Similar logic applies to the other terms, so one obtains to the desired order that 
\begin{align}
    N\langle \hat{\theta} &- \theta^* \rangle = \frac{\langle u_s u_s' \rangle}{\langle u_s' \rangle^2} - \frac{1}{2 \langle u_s' \rangle}\bigg[\frac{\langle u_s^2 \rangle \langle u_s'' \rangle}{\langle u_s' \rangle^2} + N C_{ij} \langle u_{s,ij} \rangle \nonumber \\&+ \frac{N C_{ij} \langle u_{s,i} \rangle \langle u_{s,j} \rangle \langle u_s'' \rangle}{\langle u_s' \rangle^2}
    - \frac{N C_{ij} \langle u_{s,i} \rangle \langle u'_{s,j} \rangle}{\langle u_s'\rangle} \bigg].
\end{align}
All expectation values are $O(1)$, so the right-hand side is manifestly also $O(1)$.
Since the right-hand side is evaluated at $w^*, \theta^*$, we have $\langle u_s^2 \rangle = - \langle u_s' \rangle$, so
\begin{align}
\label{eq:bc}
    \langle \hat{\theta} - \theta^* \rangle &= \frac{\langle u_s u_s' \rangle}{N\langle u_s' \rangle^2} - \frac{1}{2 N \langle u_s' \rangle}\bigg[-\frac{\langle u_s'' \rangle}{\langle u_s' \rangle}  + N C_{ij} \langle u_{s,ij} \rangle \nonumber\\
   &\:\:+ \frac{N C_{ij} \langle u_{s,i} \rangle \langle u_{s,j} \rangle \langle u_s'' \rangle}{\langle u_s' \rangle^2}- \frac{N C_{ij} \langle u_{s,i} \rangle \langle u'_{s,j} \rangle}{\langle u_s'\rangle} \bigg] \nonumber\\
    &= \frac{\langle u_s u_s' \rangle}{N \langle u_s' \rangle^2} - \frac{1}{2 \langle u_s' \rangle}\bigg[\langle u''_s \rangle \sigma^2_\text{GS}\nonumber \\
    &\qquad-\frac{ C_{ij} \langle u_{s,i} \rangle \langle u'_{s,j} \rangle}{\langle u_s'\rangle} +  C_{ij} \langle u_{s,ij} \rangle\bigg].
\end{align}
We can now define a bias corrected estimator $\hat{\theta}_{\text{BC}}$ by subtracting off the empirical estimate of \Eq{eq:bc}.

With this bias-corrected expression in hand, we return to the mixture fraction task in the Gaussian toy problem from \Sec{sec:gaussian}.
We use the Bootstrap protocol with $M\in\{8,16,32 \}$, using the same architecture and testing methodology described earlier.
In \Fig{fig:residuals_bias}, we show the distribution of naive estimate $\hat{\kappa}$ and bias-corrected version $\hat{\kappa}_\text{BC}$.
It is immediately evident that $\hat{\kappa}$ is substantially biased for $M > 8$, but that the bias correction largely mitigates this for both $M = 16$ and $M = 32$.

In \Fig{fig:bias_correction}, we show the mixture fraction coverage for these configurations.
We show the coverage both for $\hat{\kappa}$ and for $\hat{\kappa}_{\text{BC}}$, using in both cases the leading order expression for the variance given by \Eq{eq:GS}.
We see that the bias correction substantially improves the coverage for $M = 32$, where the bias becomes comparable to the spread of the estimator.

We observe empirically that the variance of  $\hat{\kappa}_\text{BC}$ is about $15\%$ larger than that of $\hat{\kappa}$ for $M=32$; the variances of these estimators are identical at $O(N^{-1})$, the order at which we truncate, so this suggests that even better coverage could be obtained by calculating the next-to-leading contribution to the variance.
This would be a substantially more involved calculation, though, so we leave this to future work.

\bibliography{biblio}

\end{document}